\documentclass[a4paper,fleqn,usenatbib]{mnras}
\usepackage{newtxtext,newtxmath}
\usepackage[T1]{fontenc}
\usepackage{ae,aecompl}
\usepackage{graphicx}	% Including figure files
\usepackage{amsmath}	% Advanced maths commands
\usepackage{amssymb}	% Extra maths symbols
\usepackage{float}
\usepackage[caption = false]{subfig}
\def\ltsima{$\; \buildrel < \over \sim \;$}
\def\simlt{\lower.5ex\hbox{\ltsima}}
\def\gtsima{$\; \buildrel > \over \sim \;$}
\def\simgt{\lower.5ex\hbox{\gtsima}}
\def\cgs{{erg cm$^{-2}$ s$^{-1}$}}
\def\ergs{{erg s$^{-1}$}}
\def\cm2{{cm$^{-2}$}}

\def\lum{{$L_{\rm X}$}}

\def\p1{{Paper I}}

\def\xmm{{\em XMM--Newton}}
\def\chandra{{\em Chandra}}

\def\swift{{\em Swift}}

\def\xmm{{\em XMM--Newton}}

\def\nh{{$N_{\rm H}$}}

\def\xray{{X--ray}}
\def\f14{{10$^{-14}$}}
\def\f13{{10$^{-13}$}}
\def\f12{{10$^{-12}$}}
\def\f11{{10$^{-11}$}}

\def\4u{{4U~1344$-$60}}

\def\lir{{$L_{\rm IR}$}}

\def\liragn{{$L_{\rm IR}^{\rm AGN}$}}
\def\lbol{{$L_{\rm Bol}$}}

\def\nus{{\em NuSTAR}}

\def\fct{$f_{\rm CT}$}
\def\f2{$f_{\rm 2}$}
\def\ecut{$E_{\rm cut}$}

\title[CT AGN at high z]{The Chandra COSMOS Legacy Survey: Compton Thick AGN at high redshift}

\author[G. Lanzuisi et al.]{G. Lanzuisi$^{1, 2}$\thanks{E-mail: giorgio.lanzuisi2@unibo.it},
F. Civano$^{3, 4}$, S. Marchesi$^{5}$, A. Comastri$^{2}$, M. Brusa$^{1, 2}$, R. Gilli$^{2}$, \newauthor
C. Vignali$^{1, 2}$, G. Zamorani$^{2}$, M. Brightman$^6$, R. E. Griffiths$^7$ and A. M. Koekemoer$^8$ \\
\\
$^1$ Dipartimento di Fisica e Astronomia, Universit\`a  di Bologna, Via Gobetti, 93/2, I-40129 Bologna, Italy\\
$^2$ INAF - Osservatorio di Astrofisica e Scienza dello Spazio di Bologna, Via Gobetti, 93/3, I--40129 Bologna, Italy \\
$^3$ Harvard-Smithsonian Center for Astrophysics, 60 Garden Street, Cambridge, MA 02138, USA \\
$^4$ Yale Center for Astronomy and Astrophysics, 260 Whitney Avenue, New Haven, CT 06520, USA \\
$^5$ Department of Physics \& Astronomy, Clemson University, Clemson, SC 29634, USA \\
$^6$ Cahill Center for Astrophysics, California Institute of Technology, 1216 East California Boulevard, Pasadena, CA 91125, USA \\
$^7$ Department of Physics \& Astronomy, University of Hawaii at Hilo, 200 W. Kawili Street, Hilo, HI 96720, USA \\
$^8$ Space Telescope Science Institute, 3700 San Martin Drive, Baltimore, MD 21218, USA
}
\date{Accepted 2018 July 25. Received 2018 July 25; in original form 2018 March 22}
\pubyear{2017}

\begin{document}
\label{firstpage}
\pagerange{\pageref{firstpage}--\pageref{lastpage}}
\maketitle

\begin{abstract}

The existence of a large population of Compton thick (CT, \nh$>10^{24}$\cm2) AGN is a key ingredient of most Cosmic X-ray background synthesis models. 
However, direct identification of these sources, especially at high redshift, is difficult, due to flux suppression and complex spectral shape produced by CT obscuration. 
We explored the \chandra\ COSMOS Legacy point source catalog, comprising 1855 sources to select, via X-ray spectroscopy, a large sample of CT 
candidates at high redshift. Adopting a physical model to reproduce the toroidal absorber and a Monte-Carlo sampling method, we selected 67 individual sources 
with $>5\%$ probability of being CT, in the redshift range $0.04\simlt z \simlt3.5$.
The sum of the probabilities above \nh$>10^{24}$ \cm2 gives a total of 41.9 {\it effective} CT AGN, corrected for classification bias. 
We derive number counts in the 2-10 keV band in three redshift bins. The observed {log}N-{log}S is consistent with an increase of the intrinsic 
CT fraction (\fct) from $\sim0.30$ to $\sim0.55$ from low to high redshift. 
When rescaled to a common luminosity ({log}(\lum/erg/s)$=44.5$) we find an increase from \fct$=0.19_{-0.06}^{+0.07}$ to \fct$=0.30_{-0.08}^{+0.10}$ 
and \fct$=0.49_{-0.11}^{+0.12}$ from low to high z. This evolution can be parametrized as 
\fct$=0.11_{-0.04}^{+0.05}(1+z)^{1.11\pm0.13}$. 
Thanks to HST-ACS deep imaging, we also find that the fraction of CT AGN in 
mergers/interacting systems increases with luminosity and redshift  
and is significantly higher than for non CT AGN hosts.
\end{abstract}


%%%%%%%%%%%%%%%%% BODY OF PAPER %%%%%%%%%%%%%%%%%%

\section{Introduction}

We have known for twenty years that a large fraction (up to 50\%) of local AGN
are obscured by large amounts of gas and dust (e.g. Risaliti et al. 1999), 
above the Compton Thick (CT) threshold\footnote{Equivalent Hydrogen column density 
\nh$\geq\sigma_T^{-1}\sim1.6\times10^{24}$ cm$^{-2}$.
At these high column densities the obscuration is mainly due to 
Compton-scattering, rather than photoelectric absorption.}.
A sizable intrinsic fraction of CT AGN (\fct) is required in most Cosmic X-ray Background (CXB) synthesis models
(e.g. Comastri et al. 1995, Gilli et al. 2007, G07 hereafter) in order to reproduce the hump observed at 20-30 keV 
in the background spectrum (e.g. Ballantyne et al. 2011).
However, the value of \fct\ derived in this way is highly uncertain, 
ranging from $\sim0.1$ to $\sim0.3-0.4$ (Treister et al. 2009, 
G07, Ueda et al. 2014) due to degeneracies between several model 
parameters, e.g. the primary continuum photon index, the reflection fraction, 
the \nh\ distribution above $10^{24}$ cm$^{-2}$, and the high energy cut-off (see e.g. Akylas et al. 2012).

X-rays are able to provide the smoking gun of CT obscuratio,
thanks to the unique spectral signatures observable, i.e. the flat continuum below $\sim10$ keV and the strong 
Fe K$\alpha$ emission line at 6.4 keV.
Furthermore, above \lum\ $\sim 10^{42}$ \ergs\ in the 2-10 keV band, the contamination by star-forming galaxies
is almost negligible.
Finally, \xray\ spectroscopy is favored by the redshift effect: going at high redshift, the Compton hump at 20-30 keV
becomes observable by \chandra\ and \xmm\, and the Fe K$\alpha$ line moves toward lower energies, where the effective area
of current \xray\ telescopes is larger.

However, collecting  large samples of CT AGN beyond the local Universe 
remains difficult, for three main reasons:\\
i) The observed fraction of CT AGN steeply rises from $\sim0$ to the intrinsic value (e.g. 0.3-0.4)
only below a certain flux (e.g. $F<<10^{-14}$ \cgs\ in the 2-10 keV band, or $10^{-13}$ \cgs\ in the 10-40 keV band, 
see e.g. G07, Ricci et al. 2015) and therefore it is mandatory to reach deep 
sensitivities over large areas, in order to collect sizable samples of CT AGN.\\
ii) For a given intrinsic luminosity, CT AGN are a factor 30-50 fainter, below 10 keV rest frame, than unobscured AGN,   
requiring long exposures to collect even a few tens of X-ray counts per source.\\ 
iii) The transition between Compton-thin and Compton-thick absorption 
(i.e. below or above \nh$>\sigma_T^{-1}\sim1.6\times10^{24}$ \cm2) is smooth (Murphy \& Yaqoob 2009, MY09 hereafter),
requiring a tailored analysis (see. e.g. Buchner et al. 2014) with the use of the full \nh\ probability 
distribution function (PDF) when selecting CT AGN, in order to avoid misclassification in one direction or the other.\\
For these reasons, even in the deepest X-ray fields, different analysis of the same samples
(e.g. Tozzi et al. 2006, Brightman \& Ueda 2012, BU12 hereafter, Georgantopoulos et al. 2013)
give results not always in agreement (see Castell\'o-Mor et al. 2013 and Liu et al. 2017).

\nus, sensitive above 10 keV, is now placing new constraints on the 
observed \fct\ at low redshift and
relatively bright fluxes, even if limited by small sample size.
Lansbury et al. (2017) find \fct$\sim0.3$ at z$\simlt0.1$ down to  $F_{\rm 10-40}\sim10^{-13}$ \cgs. 
At intermediate redshift ($z\sim0.5$), Civano et al. (2015) found an observed  
\fct$\sim0.2$, while Zappacosta et al. (2018) found
 an intrinsic \fct\ between $0.1-0.56$. Finally, Masini et al. (2018) 
derived an observed \fct$=0.11\pm0.02$ down to $F_{\rm 10-40}\sim10^{-13}$ at $z\sim1$.

Therefore, despite their expected intrinsic large fraction, CT AGN are very difficult to identify 
beyond the local Universe, resulting
in a small/negligible number of CT AGN blindly identified in medium/deep X-ray surveys 
(e.g. Tozzi et al. 2006, Comastri et al. 2011, Georgantopoulos et al. 2013, Lanzuisi et al. 2013, Civano et al. 2015, 
Marchesi et al. 2016, but see Buchner et al. 2014 and Brightman et al. 2014 for a different approach). 

Here we present the selection of 67 CT AGN candidates among the 1855 extragalactic sources with spectral analysis
from the \chandra\ COSMOS Legacy catalog (Marchesi et al. 2016, M16 hereafter).
Sec.~2 describes the sample selection and spectral analysis. In Sec.~3 we explore the relation between 
X-ray luminosity (observed and intrinsic) and IR luminosity.
Sec~4 presents the number counts of CT AGN in three redshift bins.
In Sec.~5 we derive the intrinsic \fct\ as a function of luminosity and redshift,
and in Sec.~6 we exploit the HST-ACS coverage in the COSMOS field to derive merger fraction 
for CT AGN in three luminosity bins.
Sec.~7 discusses these results.

\section{Sample selection}
\subsection{The COSMOS survey} 

The 2 deg$^2$ area of the {\it HST} COSMOS Treasury program 
is centered at 10:00:28.6, +02:12:21.0 (Scoville et al. 2007).
The field has an unrivaled deep and wide multi-wavelength coverage, 
from the optical band ({\it Hubble, Subaru, VLT}
and other ground-based telescopes), to the infrared ({\it Spitzer, Herschel}),
\xray\ (\xmm, \chandra\ and \nus) and radio bands (VLA at 1.4 and 3GHz and VLBA).
Large dedicated ground-based spectroscopic programs 
with all the major optical telescopes have been completed.
Very accurate photometric redshifts are available for both the galaxy (Ilbert et al. 2009) 
and the AGN population (Salvato et al. 2011, Marchesi et al. 2016b).

The COSMOS field has been observed in X-rays with \xmm\ for a total of $\sim1.5$ Ms at a rather homogeneous depth of
$\sim$60 ks over $\sim2$ deg$^2$ (Hasinger et al. 2007, Cappelluti et al. 2009, Brusa et al. 2010),
and by \chandra\ with deeper observations of $\sim160$ ks: the central deg$^2$ was observed 
in 2006-2007 (Elvis et al. 2009, Civano et al. 2012) for a total of 1.8 Ms, while additional 1.2 deg$^2$ were covered
more recently (2013-2014) by the Chandra COSMOS Legacy survey, for a total of 2.8 Ms (Civano et al. 2016).
56\% of the \chandra\ detected sources have a spectroscopic redshift (Marchesi et al. 2016b).

We started from the results of the spectral analysis of the full \chandra\ COSMOS Legacy catalog
performed in M16.
Their catalog contains 1855 extragalactic sources with more than 30 net counts in the full 0.5-8 keV band.
This threshold allows the derivation of basic spectral properties (\nh, \lum, see Lanzuisi et al. 2013).
For each of them a simple spectral fit was performed in M16, including a powerlaw
modified by a local neutral absorber, fixed to the Galactic value in the direction of the field, 
plus a variable neutral absorber at the source redshift. 
The power law photon index was left free to vary only for sources with more
than 70 net counts, due to the degeneracy between this parameter and \nh. 

In 67 cases a second power law was needed at 90\% confidence level (i.e. with an improvement of the fit of $\Delta Cstat>2.71$, see also Tozzi et al. 2006, Brightman et al. 2014)
in order to reproduce the emission emerging in the soft band above the obscured primary powerlaw, 
while in 141 sources an emission line was needed at the same c.l. 
to reproduce the Fe K$\alpha$ line at 6.4 keV.
Fig. 1 shows the distribution of \nh\ vs. photon index for the 1855 sources, divided on the basis of the optical/SED 
classification: red for type-2s, blue for type-1s and green for galaxies. 
This classification is based on the presence or lack of broad emission lines in the optical spectra when available, or 
on the SED best fit template class (see M16 for details).

The procedure adopted in M16 is not optimized to look for CT AGN, and indeed only 5 sources were found to be
in the CT regime.  In general, simple models such as a single powerlaw modified by photo-electric absorption 
are not able to correctly identify CT AGN, because:\\

%%%%%%%%%%%%%%%%%%%%%%%%%%%%%%%%%%%%%%%%%%%%%%%%%%%%%%%%%%%%%%%%%%%%
\begin{figure}
	\includegraphics[width=8.5cm]{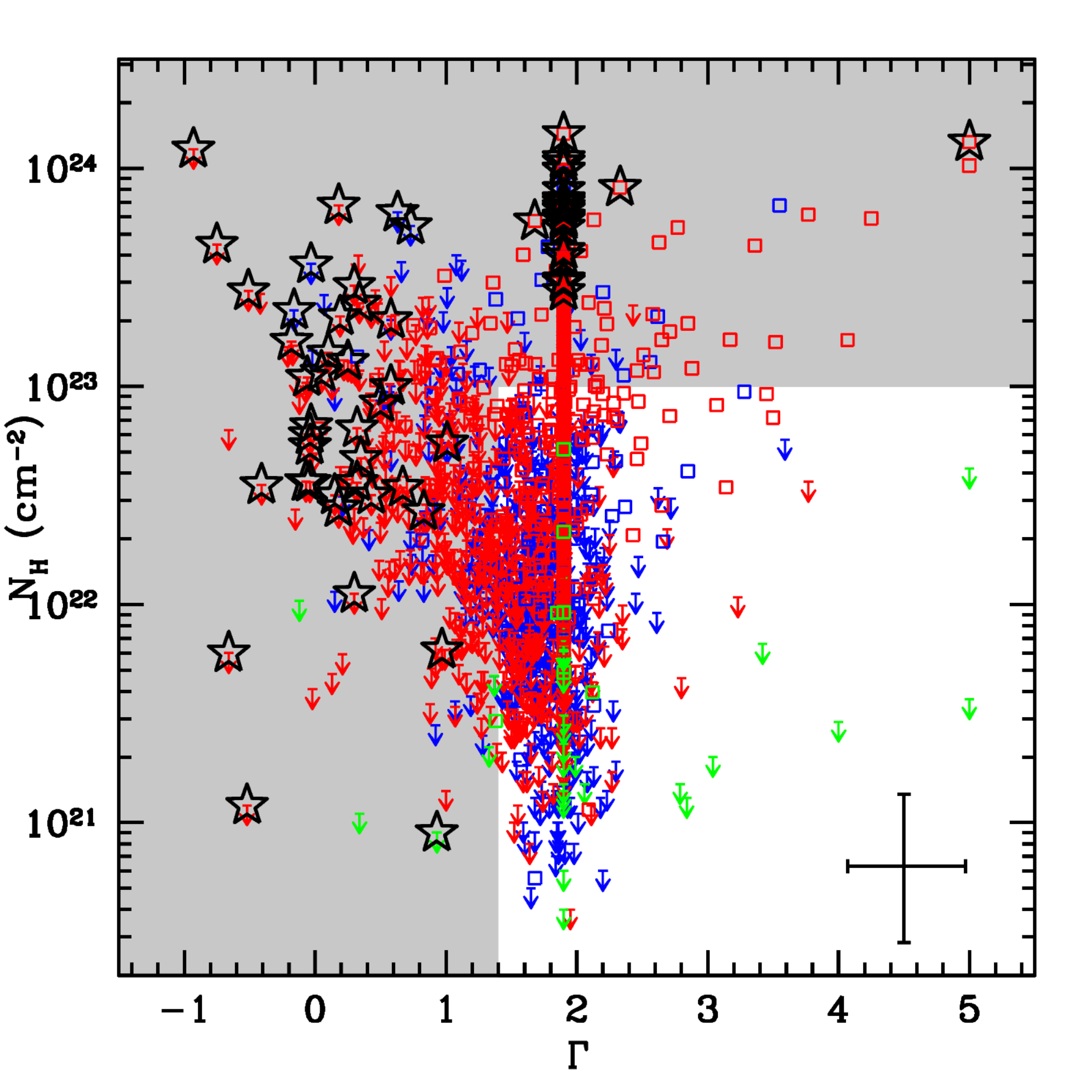}
    \caption{Distribution of photon index vs. column density derived in M16 for the 1855 extragalactic sources
    with more than 30 net counts, color-coded for optical type (blue type-1, red type-2, green galaxies).
    Squares represent \nh\ detections, arrows are 90\% upper-limits.
    The gray area shows the search region for CT candidates: sources in this area were 
    re-analyzed with the physical model described in Sec. 2.2, and the \nh\ PDF was derived. 
    The black stars highlight sources that have at least 5\% of their \nh\ PDF above the CT threshold, $10^{24}$ \cm2.
    The few sources with $\Gamma\simgt4$ are star-forming galaxies with thermal X-ray spectra, 
    or oscured sources with large errors in both \nh\ and $\Gamma$.}
    \label{fig:gam_nh}
\end{figure}
%%%%%%%%%%%%%%%%%%%%%%%%%%%%%%%%%%%%%%%%%%%%%%%%%%%%%%%%%%%%%%%%%%%%

i) neutral, photo-electric absorption components, such as {\it wabs} or similar in {\it XSPEC} (Arnaud 1996),
do not take into account Compton scattering, which becomes important above a few $\times10^{23}$ \cm2,
and do not allow the modeling of a realistic absorber geometry.\\

ii) highly obscured spectra can be well reproduced also with a flat powerlaw, 
$\Gamma\simlt1.4$ (George \& Fabian 1991, Georgantopoulos et al. 2011a), typically having a low \nh\ value.
In this case the powerlaw reproduces the flat continuum typical of reflection-dominated sources, and the 
derived \nh\ can be heavily underestimated.

Therefore sources with \nh\ above $10^{23}$ \cm2 and/or photon index below 1.4 are candidate
highly obscured sources and their X-ray spectra need to be properly modelled in order to retrieve 
a more accurate \nh\ estimate.
In the next section we describe our novel approach (see also Akylas et al. 2016) 
that combines physically motivated models, such as {\it MYtorus} 
(MY09), with Monte Carlo Markov Chain parameter estimation techniques, 
and the use of the full probaility distribution function
of the column density, to select CT AGN.

\subsection{X-ray modeling}

%%%%%%%%%%%%%%%%%%%%%%%%%%%%%%%%%%%%%%%%%%%%%%%%%%%%%%%%%%%%%%%%%%%%
\begin{figure*}
\begin{center}
\includegraphics[width=5.5cm,height=4cm]{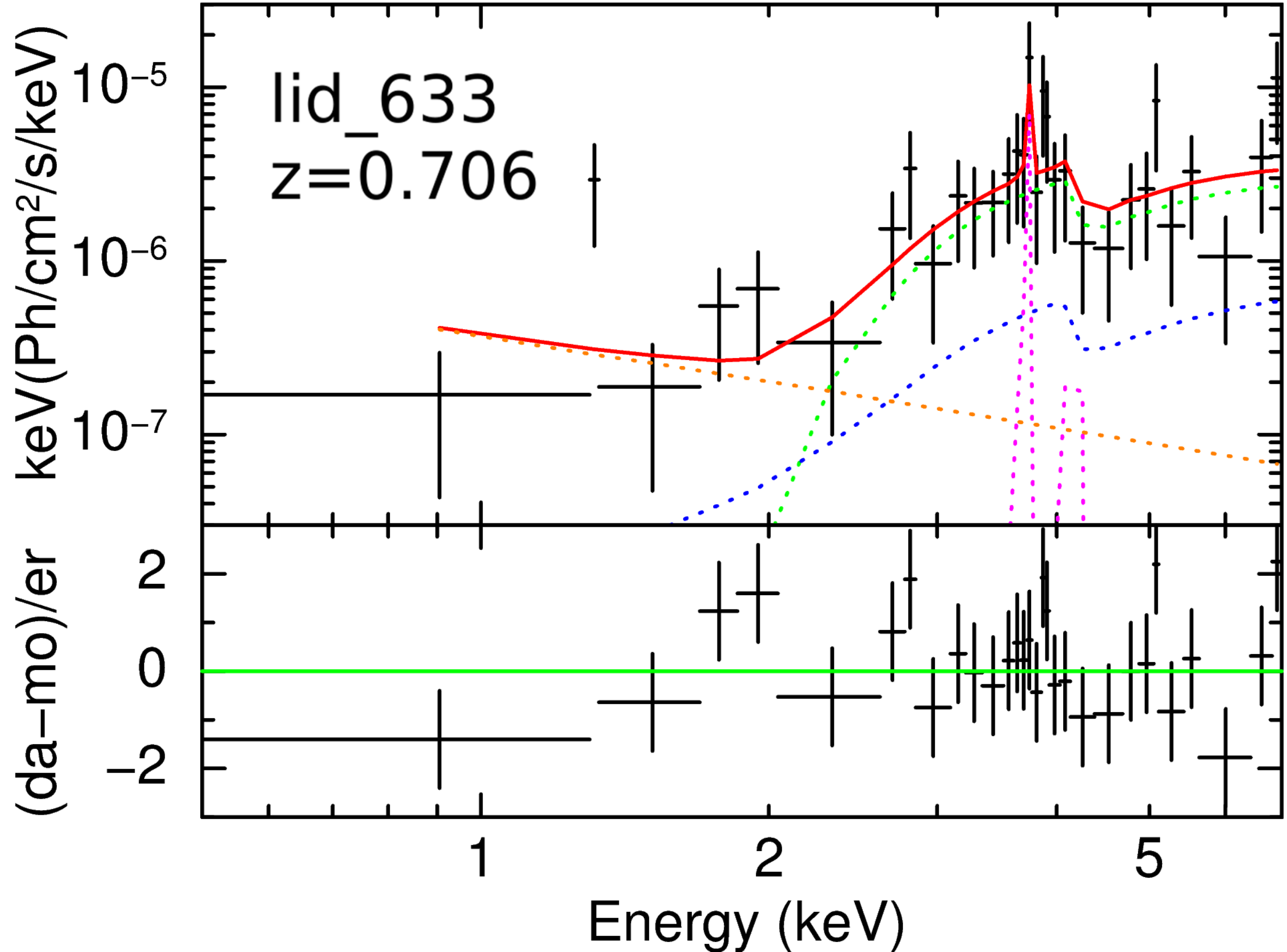}\hspace{0.2cm}\includegraphics[width=5.5cm,height=4cm]{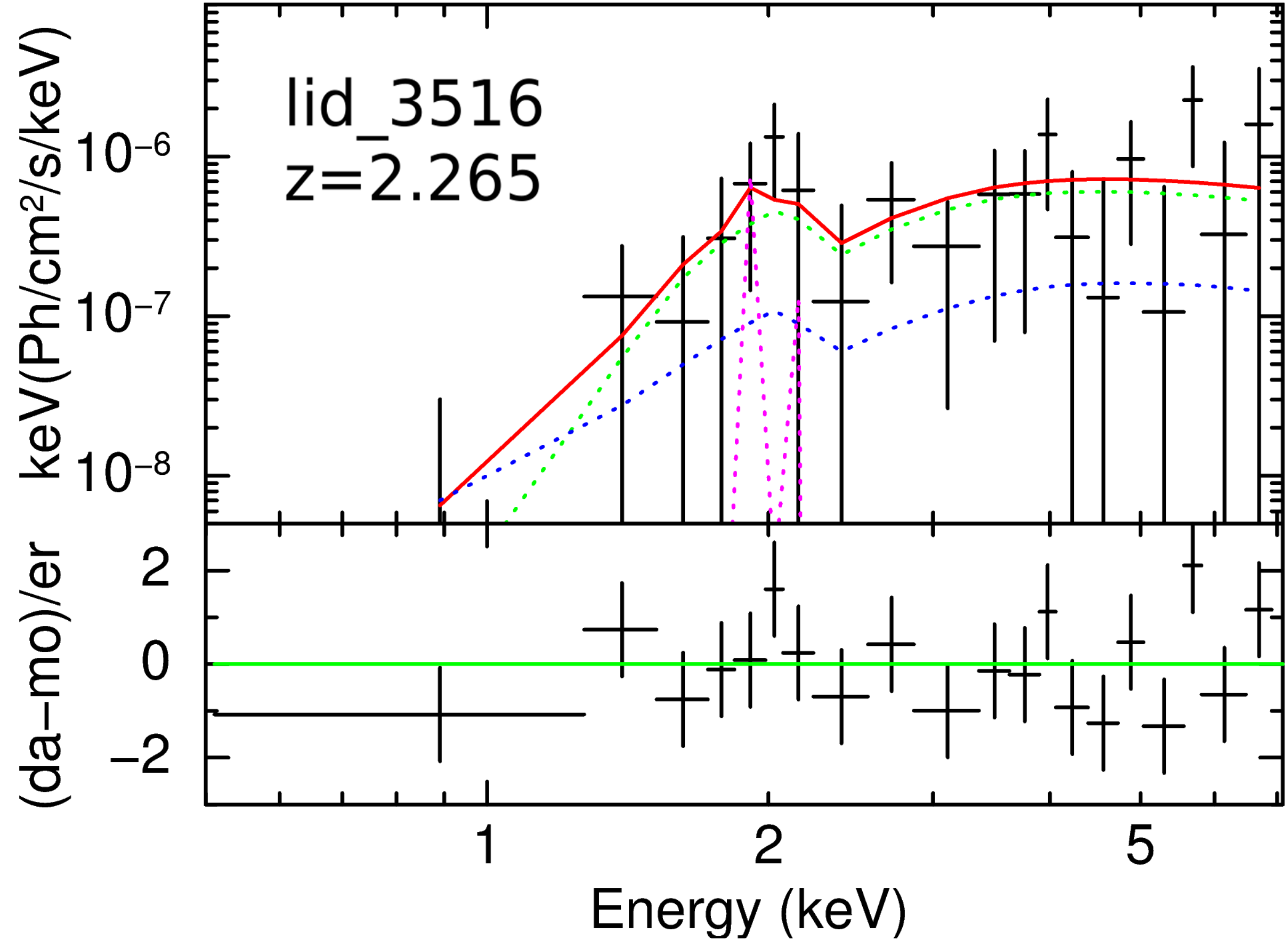}\hspace{0.2cm}\includegraphics[width=5.5cm,height=4cm]{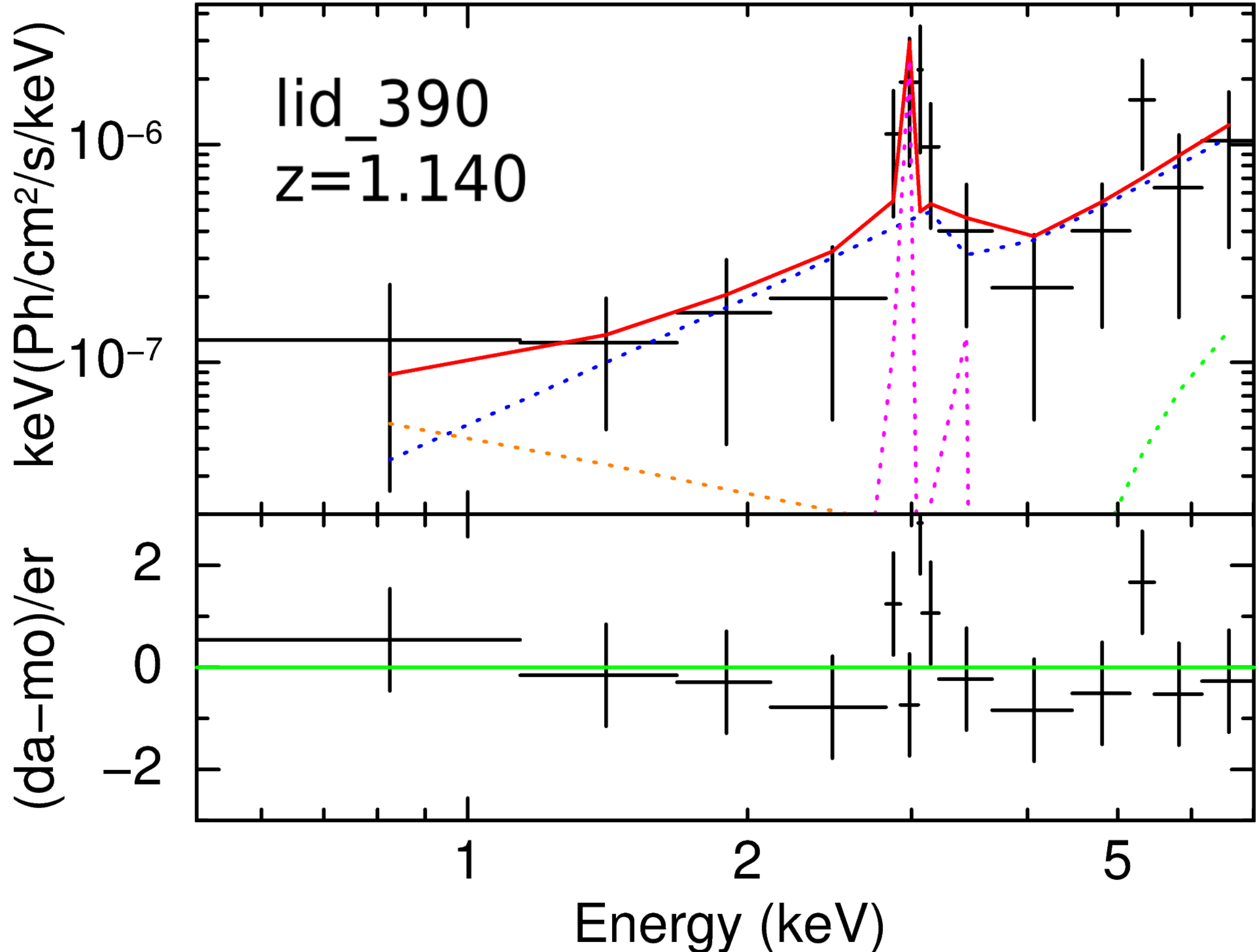}\vspace{0.3cm}
\includegraphics[width=5.5cm,height=4cm]{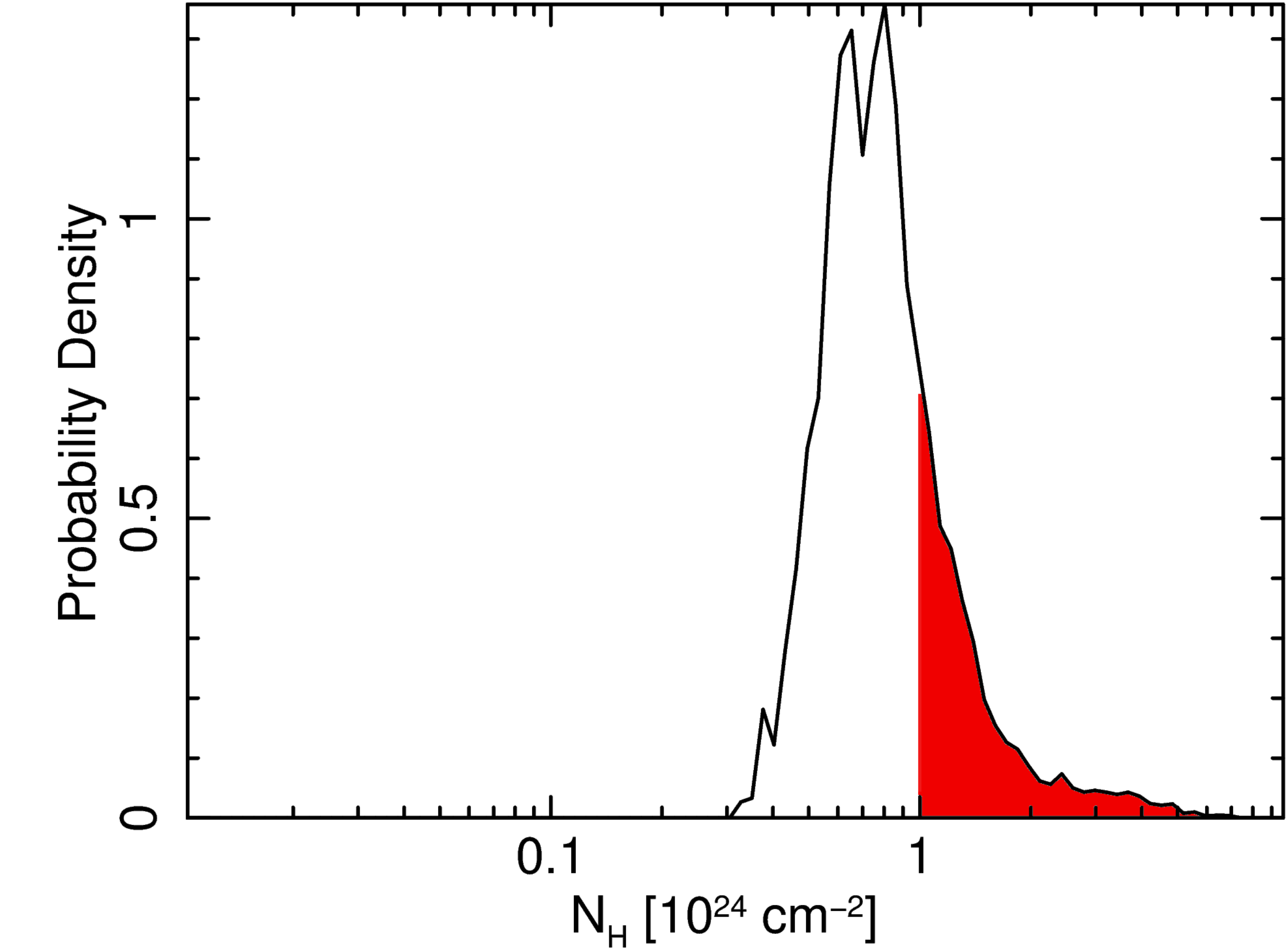}\hspace{0.2cm}\includegraphics[width=5.5cm,height=4cm]{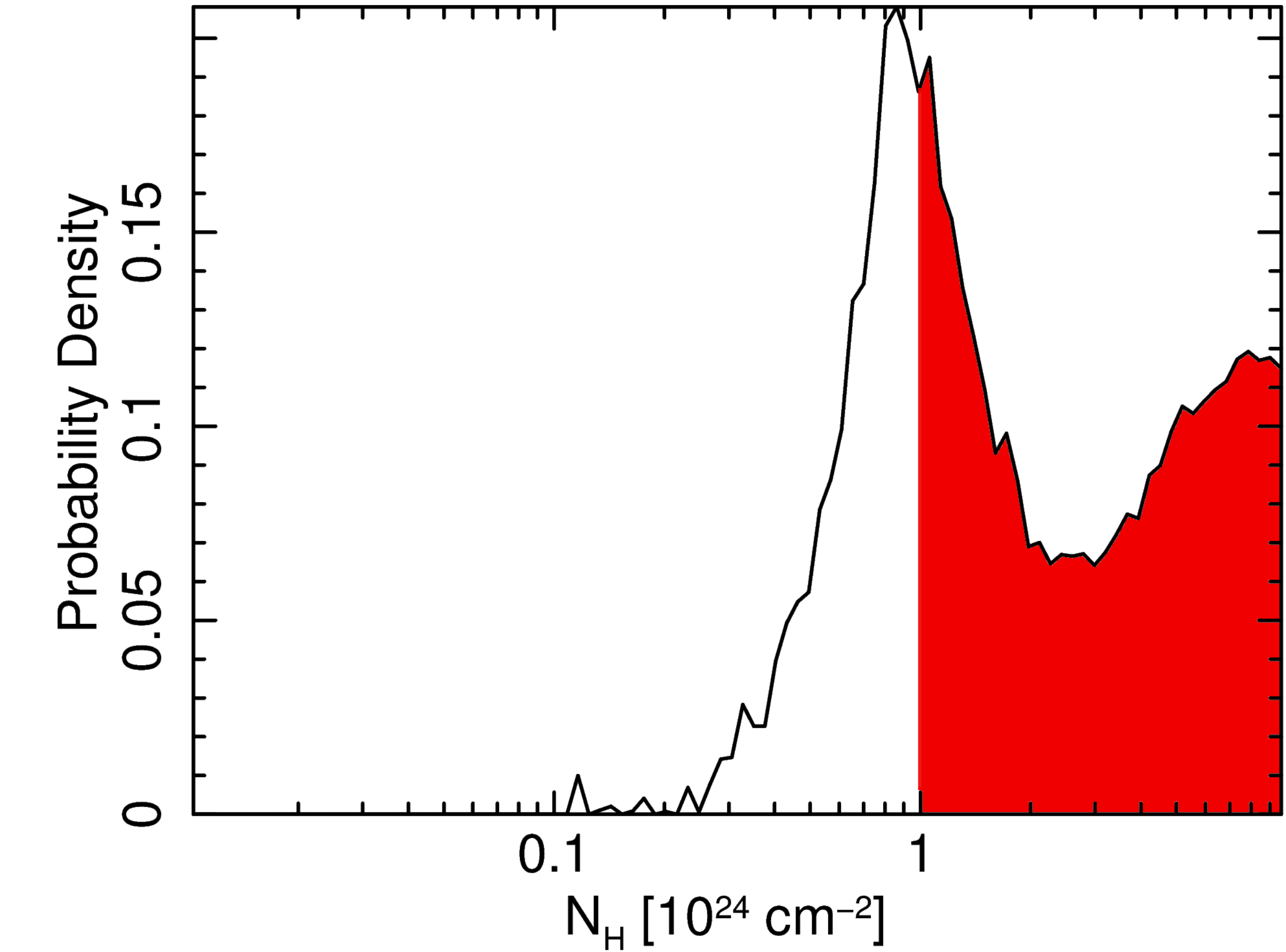}\hspace{0.2cm}\includegraphics[width=5.5cm,height=4cm]{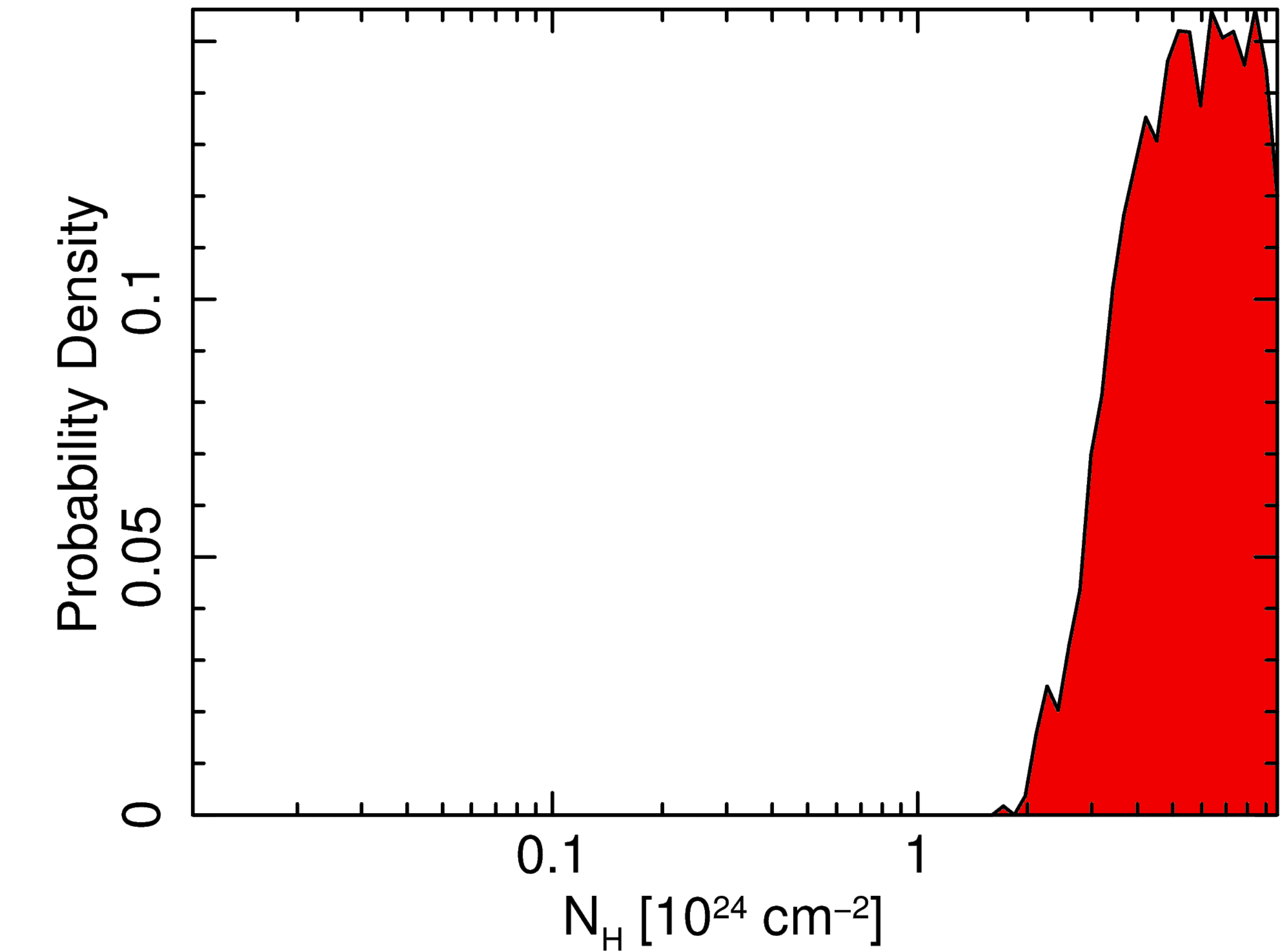}
\caption{{\it Top:} Unfolded spectrum and data-to-model ratio of the CT candidates lid\_633, lid\_3516 and lid\_390 at 
$z=0.706$,  $z=2.265$ and  $z=1.140$ respectively. 
In green we show the obscured powerlaw, in blue the reflection component, in magenta the emission lines component, in orange 
the scattered powerlaw emerging in the soft. The red curve shows the total best fit model.
{\it Bottom:} PDF of \nh\ for the three sources shown above. 
The red area shows the fraction of \nh\ PDF above the CT threshold (column 5 in Tab. 1).
}
\label{fig:spec1}
\end{center}
\end{figure*}
%%%%%%%%%%%%%%%%%%%%%%%%%%%%%%%%%%%%%%%%%%%%%%%%%%%%%%%%%%%%%%%%%%%%%%%

We reanalyzed all the 662 sources in the gray-shaded area of Fig.~\ref{fig:gam_nh}, with the physical model {\it MYtorus}
that self-consistently takes into account photoelectric absorption, Compton scattering,
cold reflection and fluorescent emission in a fixed toroidal geometry.
This model has several free parameters, and given the limited quality of the data available here, 
we decided to fix some of them.

The inclination angle between the line-of-sight and the axis
of the torus, $\Theta_{obs}$, is fixed to $75^\circ$
(where $\Theta_{obs} = 90^\circ$ corresponds to an edge-on observing angle),
to ensure that the primary continuum is
observed through the obscuring torus.
Adopting inclination angles of 65$^\circ$ or 90$^\circ$, the two extremes for the torus intercepting the line of sight,
traslates into a typical $\Delta$Log(\nh) of +0.3, and -0.1, respectively.
The powerlaw photon index is fixed to the canonical value 1.8 (Piconcelli et al. 2005), to obtain 
tighter constraints on the parameter \nh. 
A difference of $\pm0.3$ in the assumed $\Gamma$ translates into a typical $\Delta$Log(\nh) of $\pm0.1$
(see Appendix B for a more detailed analysis of the impact of 
changes in these two parameters on the derived \nh\ distribution).

The relative normalizations of the three {\it MYtorus} components, i.e. the absorbed powerlaw, the reflection and 
the emission line complex, are fixed to 1, i.e. the relative strength of the different components is fixed to the value
derived for the geometry, covering factor, and element abundances adopted in MY09, and no continuum variability is allowed (the 'coupled version' of the model).

The model is therefore very simple, as it uses a single geometry,
corresponding to a torus half opening angle of $60^\circ$, for all sources,  
and does not allow for any variation between the primary continuum 
and the reflection/emission line component, or for different values of \nh\ between the 
absorber along the line of sight and the reflecting medium, 
as would be possible in the decoupled version of the model.
This choice is forced by the very limited number of counts available for each source 
(65\% of the final CT candidate sample has less than 50 net counts). 

In addition to the {\it MYtorus} components, we added a secondary powerlaw, with the photon index fixed to 1.8,
to model the emission emerging in the soft band in most of the obscured spectra (e.g. Lanzuisi et al. 2015).  
The normalization of this component is bound to be $<5\%$ of the primary component, the
typical limit for obscured local AGN (e.g. Noguchi et al. 2010)\footnote{This secondary powerlaw must incorporate the reshift
information (e.g. zpowerlw in {\it XSPEC}) so that the normalization, defined at 1 keV rest frame, can be directly compared
with the one of {\it MYtorus}.}.

The definition of CT AGN implies a sharp threshold in Hydrogen-equivalent column density (in the literature typically assumed
\nh$>10^{24}$ \cm2, or formally \nh$>\sigma_T^{-1}\sim1.6\times10^{24}$ \cm2).
The \nh\ parameter typically has large uncertainties in faint, high redshift sources detected in deep surveys such as COSMOS (see Appendix~A).
Therefore, selecting CT sources based on \nh\ best fit value alone is subject to 
uncertainties and variation from one analysis to the other 
(see e.g. Castell\'o-Mor et al. 2013).

We adopt a different approach in our search for CT AGN.
We performed the spectral analysis with {\it XSPEC} v. 12.9.1,
using the Cash statistic (Cash 1979) with the
direct background subtraction option (W-stat, Wachter et al. 1979).
The spectra are binned to 3 counts per bin.
Once the best fit with the standard W-stat likelihood is obtained, 
we run a Monte Carlo Markov Chain (MCMC) within {\it XSPEC}, 
using the Goodman-Weare algorithm (Goodman \& Weare 2010) 
with $10^4$ steps to efficiently
explore the parameter space. The full representation of the parameter space can be then 
marginalized to look separately at each parameter distribution and derive the 
full probability distribution function (PDF).

With this method it is possible to properly account for cases in which 
the PDF has multiple peaks, as it is sometime the case for CT AGN candidates,
where two solutions are similarly allowed by the data, one at lower \nh\ and lower intrinsic flux
and one at higher \nh\ and flux (see e.g. Buchner et al. 2014).
Fig.~\ref{fig:integ} shows an example of such double-peaked PDF in the parameter space \nh\ vs. intrinsic flux, 
for source lid\_3516. The standard methods for error estimation would fail to correctly estimate the uncertainty in these cases,
ignoring one of the two solutions.

Thanks to this analysis we were able to select a sample of 67 obscured sources, at $0.04<z<3.46$,
that have at least $5\%$ of their \nh\ PDF above $10^{24}$ \cm2.
Black stars in Fig.~\ref{fig:gam_nh} show the CT candidates selected in this way. 
As hypothesized above, a large number of CT candidates have a flat powerlaw
($\Gamma<1.4$) with mild or negligible obscuration as best fit in the M16 catalog.
On the contrary, a number of sources with \nh$\simgt10^{23}$ \cm2 and steep powerlaw,
do not have any significant fraction of the \nh\ PDF above $10^{24}$ \cm2\ once the {\it MYtorus}
model with $\Gamma=1.8$ is adopted. 

%%%%%%%%%%%%%%%%%%%%%%%%%%%%%%%%%%%%%%%%%%%%%%%%%%%%%%%%%%%%%%%%%%%%%%%
\begin{table*}\scriptsize
        \centering
	\caption{Spectral properties of the CT AGN candidates.
	(1) Source ID from C16; (2) redshift (photometric ones are reported with 90\% c.l. errors);  
	(3) Net 0.5-7 keV counts; (4) Signal to noise ratio; (5) Fraction of PDF above \nh$=10^{24}$ \cm2 (corrected for classification bias); 
	(6) Best fit Log(\nh) in \cm2; (7) Scatter fraction in \%; (8) Observed 2-10 keV flux; (9) Observed 2-10 keV luminosity;
	(10) Absorption-corrected 2-10 keV luminosity; (11) Bolometric luminosity; (12) Cstat/d.o.f. of the best fit.
	Errors in columns 6 and 9 are at 90\% c.l. $^a$ Sources detected with \nus. The \nh\ value (from Zappacosta et al. 2018) 
	is in parenthesis in column (6).}
	\label{tab:fct}
	\begin{tabular}{lccccccccccc} % four columns, alignment for each
\hline
ID       &  z                                & Net C    & SNR        &  F$_{PDF}^{CT}$(Corr.)&Log(\nh) (Nus.)               & Sc. fr.   & {log}(F$_{2-10}$)  &  {log}(\lum$^{o}$) & {log}(\lum)    &  Log(\lbol) &  Cstat/d.o.f.   \\ 
(1)      & (2)                               & (3)      & (4)        &    (5)                &     (6)                      & (7)       &   (8)              &   (9)          &  (10)       &    (11)     &     (12)   \\
\hline                           
lid\_1791$^a$ & 0.044                        & 188.6    & 13.2   &  0.27 (0.27)     &  $23.91_{-0.09}^{+0.18}$ (24.0)   & 2.1      &  -13.22           & 41.25   & $42.16_{-0.42}^{+0.35}$ & 43.53 &50.7/56 \\
cid\_482$^a$  & 0.125                        & 121.3    & 10.2   &  0.79 (0.79)     &  $24.19_{-0.22}^{+0.25}$ (23.8)   & 0.4      &  -13.39           & 42.09   & $43.64_{-0.37}^{+0.42}$ & 45.5  &40.2/40 \\
cid\_460      & 0.187                        &  44.8    &  5.6   &  0.05 (0.05)     &  $23.62_{-0.14}^{+0.79}$          & 0.5      &  -13.83           & 42.04   & $42.69_{-0.35}^{+0.46}$ & 43.86 &24.2/18 \\
lid\_3635     & $0.52_{-0.01}^{+0.01}$       &  68.6    &  7.6   &  0.15 (0.15)     &  $23.85_{-0.11}^{+0.27}$          & 0.1      &  -13.62           & 43.03   & $43.94_{-0.36}^{+0.41}$ & 45.46 &14.9/20 \\
cid\_2862     & 0.551                        &  53.1    &  5.9   &  0.05 (0.05)     &  $23.63_{-0.15}^{+0.77}$          & 1.3      &  -13.84           & 42.95   & $43.37_{-0.35}^{+0.46}$ & 45.01 &15.6/22 \\
lid\_4053     & $0.64_{-0.24}^{+1.54}$       &  30.4    &  4.1   &  0.25 (0.42)     &  $23.78_{-0.48}^{+0.92}$          & 5.0      &  -14.11           & 42.8    & $43.52_{-0.40}^{+0.69}$ & 44.91 &36.6/20 \\
lid\_3816     & 0.674                        &  34.1    &  4.6   &  0.54 (0.54)     &  $23.94_{-0.11}^{+0.44}$          & 0.1      &  -13.84           & 42.92   & $43.98_{-0.68}^{+0.72}$ & 45.62 & 3.1/12 \\
lid\_1317     & $0.68_{-0.01}^{+0.01}$       &  52.5    &  6.2   &  0.68 (0.68)     &  $24.05_{-0.15}^{+0.43}$          & 1.1      &  -13.72           & 43.21   & $44.38_{-0.66}^{+0.84}$ & 45.57 &17.5/18 \\
lid\_633$^a$  & 0.706                        &  89.6    &  9.1   &  0.23 (0.23)     &  $23.82_{-0.21}^{+0.26}$ (23.7)   & 1.2      &  -13.46           & 43.49   & $44.33_{-0.33}^{+0.30}$   & 45.92 &33.9/28 \\
cid\_1019     & 0.730                        &  46.8    &  5.0   &  0.86 (1.43)     &  $24.03_{-0.19}^{+0.35}$          & 0.3      &  -14.00           & 43.0    & $44.11_{-0.66}^{+0.83}$ & 45.54 &17.8/20 \\
cid\_1254     & 0.751                        &  44.9    &  5.6   &  0.94 (0.94)     &  $24.24_{-0.11}^{+0.57}$          & 1.2      &  -13.85           & 42.97   & $44.42_{-0.21}^{+0.58}$ & 46.02 &21.6/18 \\
lid\_3487     & $0.77_{-0.02}^{+0.02}$       &  31.1    &  4.8   &  0.99 (0.99)     &  $25.00_{-0.55}^{... }$           & ...      &  -14.04           & 42.96   & $44.90_{-0.39}^{...}$   & 45.98 & 5.4/11 \\
lid\_3096     & $0.79_{-0.01}^{+0.01}$       &  33.9    &  5.1   &  0.84 (0.84)     &  $24.08_{-0.11}^{+0.46}$          & 0.1      &  -13.78           & 43.06   & $44.45_{-0.90}^{+0.75}$ & 45.99 & 6.2/11 \\
cid\_284$^a$  & 0.907                        &  52.9    &  6.4   &  0.56 (0.56)     &  $23.99_{-0.13}^{+0.28}$ (23.8)   & ...      &  -13.65           & 43.35   & $44.48_{-0.58}^{+0.51}$ & 47.19 &19.3/18 \\
lid\_1818     & 0.921                        &  66.8    &  7.4   &  0.99 (0.99)     &  $24.28_{-0.09}^{+0.58}$          & 3.6      &  -13.82           & 43.24   & $44.39_{-0.32}^{+0.73}$ & 45.05 & 9.9/23 \\
lid\_1850     & 0.947                        &  34.5    &  4.1   &  0.83 (1.38)     &  $24.06_{-0.39}^{+0.36}$          & 3.7      &  -14.18           & 42.99   & $44.05_{-0.62}^{+0.4 }$ & 45.46 &15.9/16 \\
cid\_576      & 0.972                        &  43.6    &  6.1   &  0.07 (0.07)     &  $23.75_{-0.19}^{+0.30}$          & 2.9      &  -13.90           & 43.39   & $44.12_{-0.31}^{+0.58}$ & 45.53 & 9.9/13 \\
lid\_3869     & $1.04_{-0.01}^{+0.04}$       &  31.2    &  5.0   &  0.41 (0.49)     &  $24.03_{-0.26}^{+0.30}$          & 3.3      &  -14.09           & 43.16   & $44.07_{-0.86}^{+1.32}$ & 46.05 & 4.6/10 \\
cid\_294      & 1.110                        &  95.2    &  8.9   &  0.7  (0.7 )     &  $24.05_{-0.25}^{+0.18}$          & 0.2      &  -13.78           & 43.65   & $44.72_{-0.32}^{+0.23}$ & 45.86 &31.9/31 \\
lid\_390      & 1.140                        &  39.0    &  5.8   &  1.0  (1.0)      &  $25.00_{-0.45}^{... }$           & ...      &  -13.82           & 43.27   & $45.25_{-0.34}^{...}$   & 47.15 &11.1/12 \\
lid\_665      & 1.176                        &  34.9    &  5.2   &  0.27 (0.27)     &  $23.90_{-0.27}^{+0.45}$          & 5.0      &  -13.93           & 43.29   & $44.16_{-0.48}^{+0.62}$ & 45.63 & 6.6/12 \\
cid\_886      & 1.215                        &  66.9    &  6.6   &  0.99 (0.99)     &  $24.68_{-0.17}^{... }$           & ...      &  -13.72           & 43.5    & $44.96_{-0.20}^{+...}$  & 46.59 &41.9/28 \\
lid\_4377     & $1.25_{-0.02}^{+0.13}$       &  34.9    &  4.2   &  0.19 (0.22)     &  $23.53_{-0.23}^{+0.74}$          & ...      &  -14.02           & 43.4    & $44.02_{-0.30}^{+0.85}$ & 45.83 &23.8/16 \\
cid\_1135     & $1.44_{-0.05}^{+0.17}$       &  41.0    &  4.7   &  0.43 (0.51)     &  $23.95_{-0.22}^{+0.44}$          & 3.0      &  -14.07           & 43.44   & $44.38_{-0.63}^{+0.73}$ & 45.98 &16.1/18 \\
lid\_3056     & $1.44_{-0.04}^{+0.02}$       &  40.5    &  5.4   &  0.81 (0.81)     &  $24.13_{-0.12}^{+0.23}$          & 0.9      &  -13.84           & 43.46   & $44.74_{-0.62}^{+0.61}$ & 46.54 &18.1/16 \\
cid\_1078     & 1.478                        &  35.2    &  4.3   &  0.4  (0.4 )     &  $23.79_{-0.11}^{+0.48}$          & ...      &  -13.96           & 43.48   & $44.31_{-0.40}^{+0.49}$ & 45.99 & 8.3/16 \\
cid\_2856     & $1.51_{-0.05}^{+0.38}$       &  35.6    &  4.5   &  0.9  (1.08)     &  $24.29_{-0.15}^{+0.53}$          & 2.9      &  -14.12           & 43.33   & $44.62_{-0.22}^{+0.53}$ & 46.35 &22.1/14 \\
cid\_1474     & 1.551                        &  36.4    &  4.5   &  0.75 (0.75)     &  $24.55_{-0.19}^{+0.41}$          & ...      &  -13.79           & 43.36   & $45.33_{-0.45}^{+0.65}$ & 47.01 &14.7/19 \\
cid\_1125     & 1.555                        &  36.7    &  4.3   &  0.29 (0.29)     &  $23.84_{-0.36}^{+0.45}$          & 2.4      &  -13.81           & 43.58   & $44.34_{-0.48}^{+0.48}$ & 47.6  &19.9/17 \\
lid\_1549     & 1.650                        &  32.6    &  4.7   &  0.35 (0.46)     &  $23.86_{-0.22}^{+0.48}$          & 2.8      &  -14.16           & 43.52   & $44.36_{-0.45}^{+0.92}$ & 45.91 &17.5/11 \\
cid\_1226     & $1.71_{-0.25}^{+0.29}$       &  64.1    &  5.5   &  0.3  (0.38)     &  $23.80_{-0.19}^{+0.46}$          & 2.7      &  -14.03           & 43.67   & $44.44_{-0.51}^{+0.54}$ & 46.12 &37.6/33 \\
lid\_3289     & 1.728                        &  30.0    &  4.1   &  0.85 (0.85)     &  $24.36_{-0.10}^{+0.35}$          & ...      &  -13.89           & 43.31   & $45.03_{-0.73}^{+0.75}$ & 46.8  & 8.1/12 \\
cid\_973      & $1.75_{-0.02}^{+0.53}$       &  39.9    &  4.6   &  0.97 (0.97)     &  $24.60_{-0.14}^{+0.36}$          & 0.5      &  -14.01           & 43.39   & $45.21_{-0.45}^{+0.47}$ & 47.1  &15.8/19 \\
cid\_370      & 1.757                        &  35.6    &  4.1   &  0.89 (0.89)     &  $24.32_{-0.17}^{+0.51}$          & ...      &  -13.92           & 43.36   & $44.93_{-0.51}^{+0.72}$ & 46.73 &17.7/21 \\
lid\_603      & 1.776                        &  43.4    &  5.3   &  0.52 (0.62)     &  $24.33_{-0.22}^{+0.56}$          & 2.9      &  -14.13           & 43.68   & $44.83_{-0.38}^{+0.57}$ & 45.96 &18.8/17 \\
cid\_713      & 1.778                        &  46.0    &  6.0   &  0.21 (0.21)     &  $23.85_{-0.11}^{+0.97}$          & 0.3      &  -13.96           & 43.69   & $44.59_{-0.29}^{+0.30}$ & 46.32 &22.7/15 \\
cid\_3234     & $1.80_{-0.07}^{+0.09}$       &  38.0    &  4.1   &  0.93 (0.93)     &  $24.43_{-0.08}^{+0.54}$          & 0.3      &  -13.91           & 43.37   & $45.13_{-0.39}^{+0.73}$ & 46.96 &16.9/21 \\
cid\_102      & 1.847                        &  73.8    &  7.2   &  0.16 (0.16)     &  $23.83_{-0.10}^{+0.41}$          & 0.8      &  -13.99           & 43.82   & $44.67_{-0.22}^{+0.30}$ & 46.18 &26.3/22 \\
cid\_1060     & $1.85_{-0.06}^{+0.07}$       &  79.8    &  6.9   &  0.15 (0.15)     &  $23.85_{-0.14}^{+0.15}$          & ...      &  -13.76           & 43.83   & $44.56_{-0.22}^{+0.27}$ & 46.7  & 6.5/10 \\
cid\_1271     & $1.97_{-0.25}^{+0.09}$       &  35.0    &  4.3   &  0.56 (0.67)     &  $23.92_{-0.13}^{+0.69}$          & 1.5      &  -14.1            & 43.64   & $44.61_{-0.43}^{+0.52}$ & 46.28 &21.0/17 \\
lid\_771      & $1.98_{-0.13}^{+0.01}$       &  39.1    &  6.1   &  0.57 (0.57)     &  $24.06_{-0.23}^{+0.24}$          & 1.1      &  -13.75           & 43.85   & $45.03_{-0.37}^{+0.39}$ & 46.81 &10.3/10 \\
lid\_3178     & $2.00_{-0.20}^{+0.17}$       &  32.4    &  5.0   &  0.71 (0.85)     &  $24.06_{-0.03}^{+0.89}$          & ...      &  -14.00           & 43.56   & $44.83_{-0.47}^{+0.41}$ & 46.54 & 6.7/11 \\
lid\_1026     & 2.003                        &  52.7    &  6.8   &  1.0  (1.0 )     &  $25.00_{-0.34}^{... }$           & 1.3      &  -13.83           & 43.97   & $45.25_{-0.39}^{...}$   & 47.21 &24.4/17 \\
cid\_1054     & $2.02_{-0.94}^{+0.05}$       &  45.1    &  5.2   &  0.1  (0.1 )     &  $23.67_{-0.18}^{+0.87}$          & 2.9      &  -14.12           & 43.83   & $44.45_{-0.30}^{+0.42}$ & 46.14 &29.3/20 \\
cid\_862      & $2.06_{-0.01}^{+0.01}$       &  60.4    &  4.7   &  0.68 (0.68)     &  $24.13_{-0.13}^{+0.44}$          & 0.5      &  -13.93           & 43.67   & $45.01_{-0.47}^{+0.56}$ & 45.72 &22.4/24 \\
cid\_3284     & $2.09_{-0.29}^{+0.25}$       &  36.4    &  4.2   &  0.59 (0.59)     &  $24.24_{-0.13}^{+0.66}$          & 0.7      &  -14.00           & 43.48   & $44.99_{-0.38}^{+0.43}$ & 46.77 &17.5/16 \\
cid\_1956     & 2.160                        &  44.6    &  4.3   &  0.98 (0.98)     &  $25.00_{-0.30}^{... }$           & ...      &  -14.01           & 43.65   & $45.30_{-0.45}^{...}$   & 47.18 &17.1/25 \\
cid\_1286     & 2.200                        &  36.6    &  5.2   &  0.75 (1.07)     &  $24.34_{-0.05}^{+0.62}$          & 3.6      &  -14.16           & 43.63   & $44.89_{-0.21}^{+0.78}$ & 46.71 & 8.4/13 \\
lid\_3516     & 2.265                        &  34.8    &  4.0   &  0.64 (0.64)     &  $23.96_{-0.29}^{+0.52}$          & ...      &  -14.12           & 43.63   & $44.72_{-0.48}^{+0.57}$ & 46.46 & 9.5/17 \\
cid\_1615     & $2.29_{-0.37}^{+0.13}$       &  78.9    &  8.1   &  0.83 (0.83)     &  $24.07_{-0.12}^{+0.21}$          & 0.4      &  -13.65           & 44.02   & $45.31_{-0.30}^{+0.39}$ & 46.5  &29.9/25 \\
cid\_1143     & 2.335                        &  51.3    &  5.0   &  0.4  (0.4 )     &  $23.88_{-0.30}^{+0.77}$          & 1.8      &  -14.15           & 43.79   & $44.69_{-0.36}^{+0.43}$ & 46.38 &41.5/28 \\
cid\_976      & 2.478                        &  59.7    &  6.0   &  0.99 (0.99)     &  $24.70_{-0.20}^{... }$           & 5.0      &  -14.08           & 43.96   & $45.35_{-0.27}^{...}$   & 47.34 &20.3/20 \\
cid\_1503     & 2.497                        &  39.5    &  4.7   &  0.54 (0.54)     &  $23.95_{-0.09}^{+0.52}$          & ...      &  -14.05           & 43.79   & $44.86_{-0.38}^{+0.45}$ & 46.67 &18.6/17 \\
cid\_610      & 2.571                        &  39.5    &  5.6   &  0.5  (0.5 )     &  $24.18_{-0.28}^{+0.51}$          & 4.8      &  -14.07           & 43.78   & $44.98_{-0.74}^{+0.72}$ & 46.84 &16.3/14 \\
cid\_360      & $2.58_{-0.05}^{+0.24}$       &  59.8    &  6.5   &  0.13 (0.13)     &  $23.82_{-0.14}^{+0.37}$          & 1.2      &  -14.07           & 43.94   & $44.79_{-0.26}^{+0.28}$ & 46.59 &22.7/21 \\
lid\_1002     & 2.612                        &  65.6    &  7.5   &  0.67 (0.67)     &  $23.98_{-0.15}^{+0.16}$          & ...      &  -13.73           & 44.07   & $45.19_{-0.24}^{+0.26}$ & 47.2  &19.7/21 \\
cid\_708      & 2.612                        &  49.5    &  6.1   &  1.0  (1.0 )     &  $25.00_{-0.22}^{... }$           & 0.3      &  -14.03           & 43.85   & $45.49_{-0.22}^{...}$   & 47.27 &21.1/18 \\
lid\_1838     & 2.624                        &  45.8    &  6.2   &  0.55 (0.55)     &  $23.97_{-0.16}^{+0.40}$          & 1.9      &  -14.10           & 43.87   & $44.84_{-0.31}^{+0.53}$ & 46.62 &18.0/14 \\
cid\_747      & 2.709                        &  50.6    &  5.6   &  0.62 (0.62)     &  $24.05_{-0.13}^{+0.47}$          & ...      &  -14.01           & 43.81   & $45.00_{-0.39}^{+0.35}$ & 46.84 &23.3/23 \\
lid\_1816     & $2.81_{-0.03}^{+0.02}$       & 114.4    & 10.3   &  0.48 (0.48)     &  $23.96_{-0.17}^{+0.19}$          & 5.0      &  -13.77           & 44.4    & $45.17_{-0.27}^{+0.28}$ & 47.16 &36.2/33 \\
cid\_2177     & $2.89_{-1.24}^{+0.45}$       &  49.0    &  5.4   &  0.39 (0.39)     &  $23.93_{-0.16}^{+0.49}$          & ...      &  -14.11           & 43.86   & $44.90_{-0.37}^{+0.37}$ & 46.73 &21.6/22 \\
cid\_45       & 2.909                        &  58.8    &  7.1   &  0.18 (0.18)     &  $23.84_{-0.16}^{+0.54}$          & 2.6      &  -14.08           & 44.07   & $44.87_{-0.27}^{+0.26}$ & 46.71 &18.4/19 \\
lid\_439      & $2.93_{-0.07}^{+0.08}$       &  72.0    &  7.3   &  0.75 (0.75)     &  $24.45_{-0.24}^{+0.48}$          & 1.3      &  -13.75           & 44.15   & $45.71_{-0.35}^{+0.38}$ & 47.94 &32.5/28 \\
cid\_965      & 3.178                        &  35.1    &  5.3   &  0.79 (0.79)     &  $24.47_{-0.20}^{+0.45}$          & 1.5      &  -14.11           & 43.71   & $45.37_{-0.57}^{+0.47}$ & 47.23 &23.3/11 \\
cid\_700      & 3.191                        &  55.3    &  6.4   &  0.66 (0.66)     &  $24.13_{-0.22}^{+0.57}$          & 2.5      &  -14.12           & 43.96   & $45.11_{-0.47}^{+0.44}$ & 46.93 &27.4/20 \\
lid\_1705     & $3.46_{-0.10}^{+0.05}$       &  44.0    &  5.7   &  0.75 (0.75)     &  $24.19_{-0.10}^{+0.65}$          & 0.1      &  -14.07           & 43.95   & $45.36_{-0.53}^{+0.45}$ & 47.18 &23.0/17 \\
lid\_283      & 3.465                        &  57.9    &  6.8   &  0.73 (0.73)     &  $24.45_{-0.25}^{+0.43}$          & 3.4      &  -14.00           & 44.11   & $45.48_{-0.50}^{+0.43}$ & 47.59 &29.3/19 \\
\hline
\label{tab:tab1}
\end{tabular}
\end{table*}
%%%%%%%%%%%%%%%%%%%%%%%%%%%%%%%%%%%%%%%%%%%%%%%%%%%%%%%%%%%%%%%%%%%%%%%

%%%%%%%%%%%%%%%%%%%%%%%%%%%%%%%%%%%%%%%%%%%%%%%%%%%%%%%%%%%%%%%%%%%%
\begin{figure*}
\begin{center}
\includegraphics[width=8.3cm,height=7.5cm]{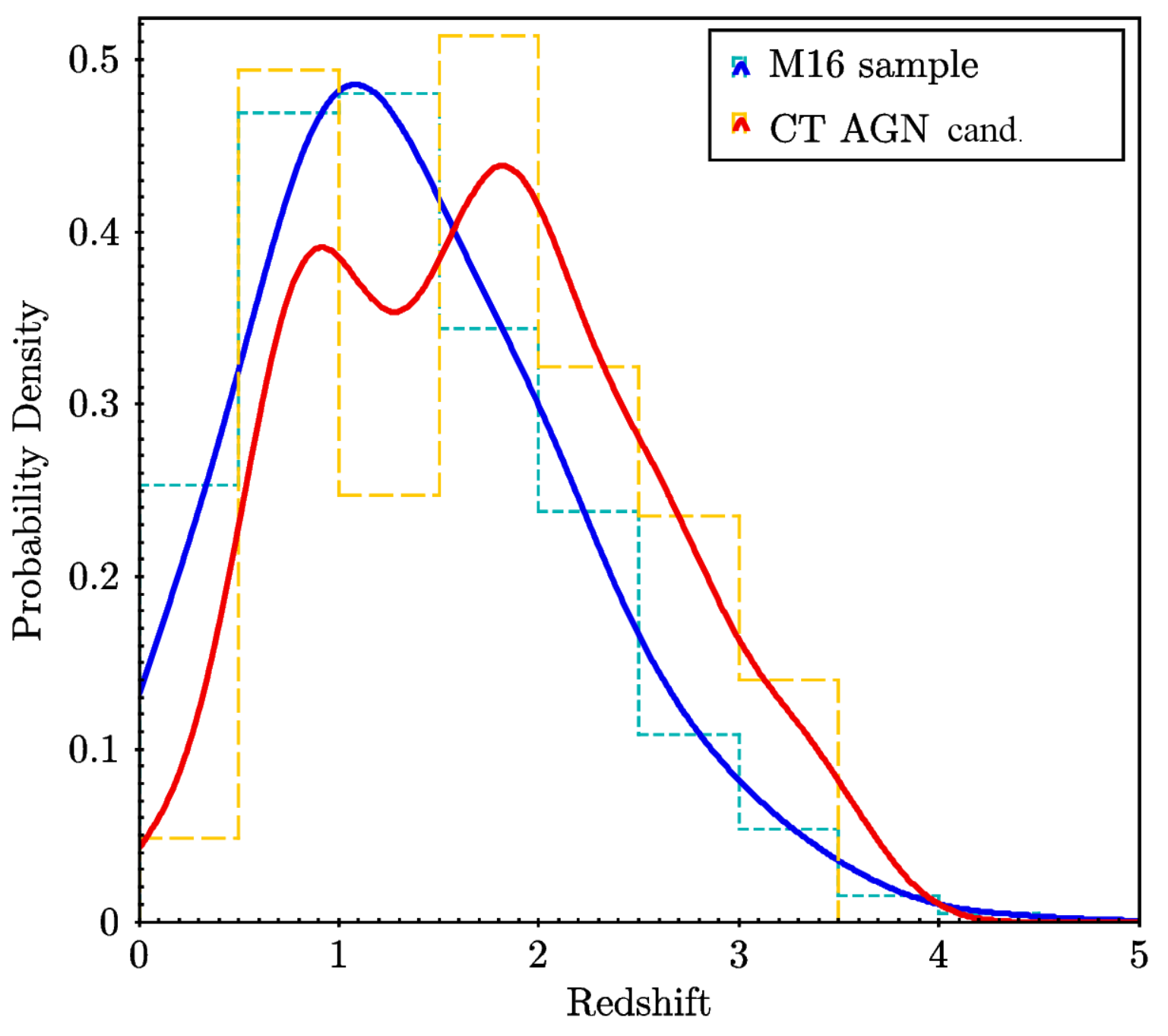}\hspace{0.5cm}\includegraphics[width=8.3cm,height=7.5cm]{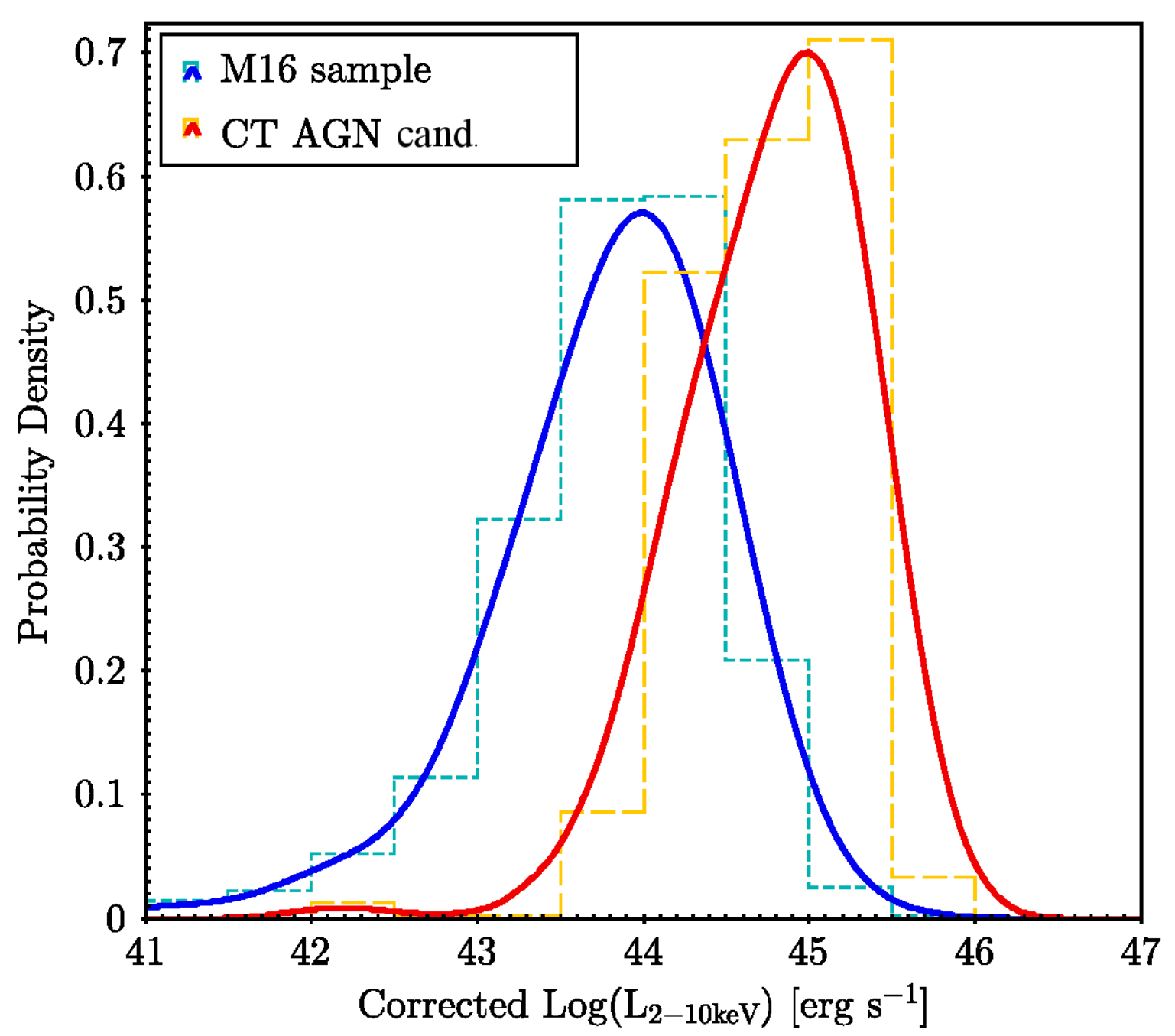}
\caption{
Redshift (left) and absorption-corrected 2-10 keV band Luminosity (right) normalized distributions for the 
CT AGN candidates (red) and the M16 sample (blue). The CT sample is corrected for classification bias and survey sensitivity,
and each source is counted only for the fraction of the PDF above the CT threshold.
The thick lines show the distribution derived with a kernel density estimation
(KDE) using a Gaussian kernel with a bandwidth of 0.3 applied to the underlying distributions (dashed histograms).
The CT AGN tend to have higher redshift with respect to the M16 sample, while the typical 
absorption-corrected 2-10 keV luminosity is one order of magnitude higher (see text).
}
\label{fig:isto_z_lum}
\end{center}
\end{figure*}
%%%%%%%%%%%%%%%%%%%%%%%%%%%%%%%%%%%%%%%%%%%%%%%%%%%%%%%%%%%%%%%%%%%%%%%

Fig.~\ref{fig:spec1} shows the unfolded spectra plus residuals and the \nh\ PDF for three CT 
candidates selected in this way 
(Fig.~D1 available in the online journal, shows spectra and \nh\ PDF for the entire sample).
the left panel shows a C-thin source with a small PDF fraction above the CT threshold, 
the central panel shows a double-peaked PDF with one solution in the C-thin regime and one in the heavily CT regime 
(see also appendix A1), while the right panel shows a heavily CT, reflection-dominated source. The upper boundary of 
\nh\ at $10^{25}$ \cm2 is 
set by the limit of the model.

If only the part of the PDF above the CT threshold (red area) is taken into account when computing the number of CT,
it is possible to construct an {\it effective} sample of CT AGN candidates, by counting
each source only for the fraction of the PDF that exceeds the CT boundary.
This means that virtually none of these sources has 100\% probability of being CT or Compton-thin, but
we can consider the sum of the probabilities  above the CT threshold for the whole sample
as a good approximation of the total number of CT in that sample.
Summing up only the fraction of the PDF of each source that is above $10^{24}$ \cm2,
we obtained a number of CT sources of N$_{\rm CT}=38.5$.
This number is stable with respect to the threshold adopted to select CT AGN, as it would be
38.1 or 38.8 if this threshold is taken at 10\% or 1\% of the PDF above $10^{24}$ \cm2, respectively.

Tab.~\ref{tab:tab1} summarizes the results for all the 67 CT AGN candidates.
All our sources have a $0.5-7$ keV signal-to-noise ratio $SNR>4$, despite 
having tipically a low number of counts (column 3), thanks to the very low background levels of \chandra.
The sources identified with cid\_ are detected in the catalog of Elvis et al. (2009) and Civano et al. (2012),
while the ones identified with lid\_ come from the catalog of Civano et al. (2016) and Marchesi et al. (2016b).
For 60\% of the sample a spectroscopic redshift is available. The remaining sources have photometric redshift, 
with good accuracy. 
The errors reported in column 2 are the 90\% c.l. errors computed from the photometric redshift PDF.
The vast majority of them are $<0.10$, a few up to $\sim0.5$ and only three sources (lid\_4053,cid\_1054 and cid\_2177) 
have errors larger than 0.9, due to the presence of secondary peaks of the PDF distribution.

We note that 4 sources at $z<1$ (labelled with $^a$ in Tab.~1) 
are detected also by the \nus\ survey performed in COSMOS (Civano et al. 2015).
The best fit \nh\ obtained by fitting \xmm\ and/or \chandra\ data 
together with \nus\  (Zappacosta et al. 2018) is consistent, within the errors, with the one derived here ,
even if there is a tendency for slightly lower \nh\ obtained when \nus\ is included in the fit (see Marchesi et al. 2018
for a systematic study of this effect on a sample of local CT AGN).

\section{\lum\ vs. \lir}

One of the key parameters that can be derived from the X-ray spectral fitting of CT AGN is the intrinsic, 
absorption-corrected 2-10 keV Luminosity (\lum).
To compute its value and realistic errors we used the {\it cflux} component
in {\it XSPEC}, applied to the unobscured, primary powerlaw\footnote{The reflection 
and emission line components have their own {\it cflux} applied, and their flux ratios 
with respect to the unabsorbed primary powerlaw are fixed to the value derived 
from the {\it MYtorus} model itself, i.e. 2.7\% and 0.6\% respectively, in the rest frame 2-10 
keV band, in order to keep the proper relative normalization between the different components.}.
%[plus a second {\it cflux}
%component applied to the reflection and emission line components, with fixed flux ratio of 3.5\%, a value
%derived by the {\it mytorus} model with \nh$=10^{24}$ \cm2\ and the other parameters as described in sec. 2.2. Too complicated?].
In this way the intrinsic flux due to the primary powerlaw becomes a free parameter of the fit,
and its errors can be evaluated self-consistently with the errors on \nh, with the same MCMC approach.
These values are then converted into \lum\ based on the luminosity distance of each source.

The sample of selected CT candidates spans a wide range in redshift, $0.04<z<3.46$
and absorption corrected X-ray luminosity $42<${log}(\lum)$<45.7$ \ergs.
Fig.~\ref{fig:isto_z_lum} (left) shows the normalized redshift distribution for the 
CT AGN sample (orange) and the M16 sample (cyan). 
The thick lines show the distribution derived with a kernel density estimation
(KDE) using a Gaussian kernel with a bandwidth of 0.3 applied to the underlying distributions for better visualization.
The CT AGN tend to have a higher redshift (mean redshift 1.71, median 1.75) with respect to the M16 sample 
(mean 1.42, median 1.29). This is a result of the positive k-correction for CT AGN in X-ray mentioned in Sec.~1. 
Fig.~\ref{fig:isto_z_lum} (right) shows the same for the absorption-corrected 2-10 keV luminosity. In this case
the CT sample has one order of magnitude higher \lum\ (mean {log}(\lum) 44.7 vs. 43.8 \ergs, median 44.8 vs. 43.9)
with respect to the M16 sample. In this case the difference is larger thanks to the 
larger correction for absorption applied in the CT sample with respect to the full catalog.

%%%%%%%%%%%%%%%%%%%%%%%%%%%%%%%%%%%%%%%%%%%%%%%%%%%%%%%%%%%%%%%%%%%%
\begin{figure*}
\begin{center}
\includegraphics[width=\columnwidth]{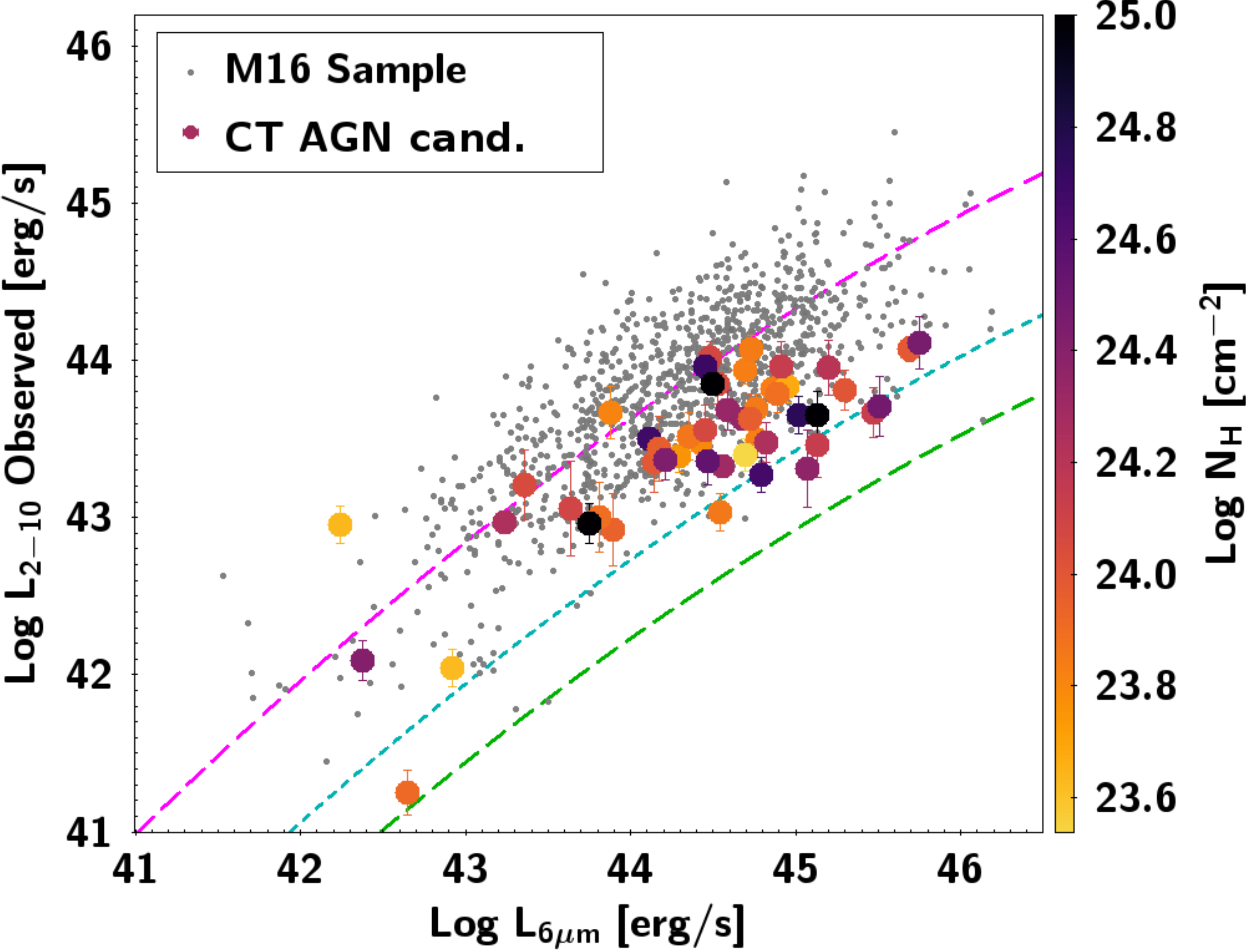}\hspace{0.5cm}\includegraphics[width=\columnwidth]{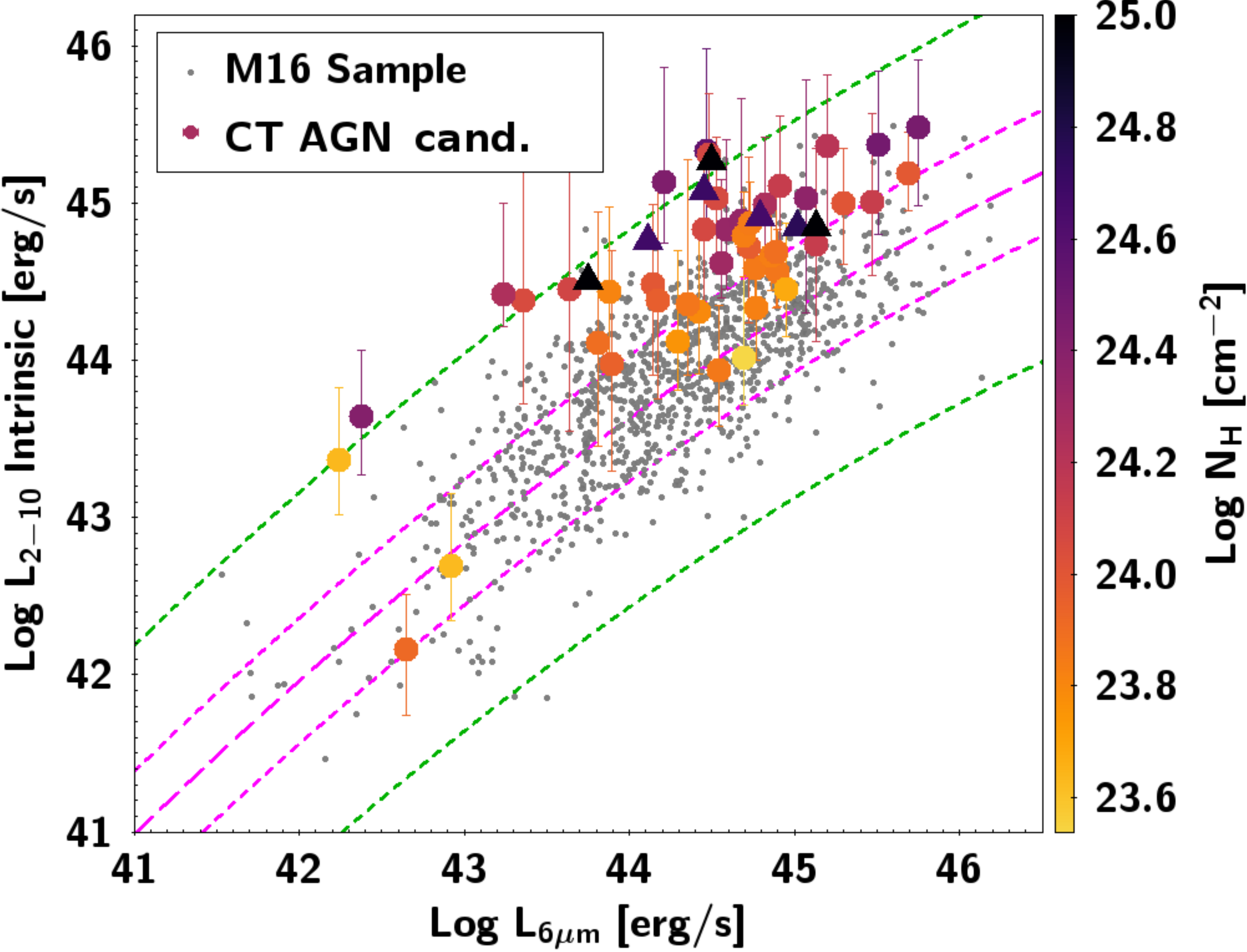}
\caption{
Distribution of the AGN Mid-IR ($6 \mu m$) luminosity, as derived from SED fitting, 
vs. 2-10 keV luminosity, 
for the 1048 sources in the COSMOS Legacy catalog with \liragn\ available (gray dots). 
Left panel shows the observed \lum\, while right panel shows 
the absorption-corrected \lum. 
Data points for the CT candidates are color-coded on the basis of Log\nh. 
Triangles show lower limits of the intrinsic 
\lum\ for sources  with a lower limit in the \nh.
The magenta dashed curve in both panels shows the relation from Stern (2015).
In the left panel the cyan dotted (green dashed) line shows the \lum\ decrement  expected for log\nh=24 (log\nh=25) \cm2,
assuming the model described in Sec.~2.2.
In the right panel, the dotted magenta (green) curves show the $1\sigma$ ($3\sigma$)
dispersion of the COSMOS Legacy sample.}
\label{fig:lumlum}
\end{center}
\end{figure*}
%%%%%%%%%%%%%%%%%%%%%%%%%%%%%%%%%%%%%%%%%%%%%%%%%%%%%%%%%%%%%%%%%%%%%%%

Indeed, given the range of corrections applied for the obscuration, 
the majority of the sources in the CT sample are in the quasar 
regime (\lum$>10^{44}$ \ergs). 
In order to verify if these luminosities are consistent with other information available for this sample, 
i.e. if our estimates of the obscuration are correct, 
we compared the intrinsic X-ray luminosity with the mid-IR ($6\mu m$) AGN luminosity, as computed 
from the spectral energy distribution (SED) 
fitting from Delvecchio et al. (2015), or Suh et al. (2017) if the source is not FIR detected, 
after removing the host star-formation emission in the same band. 
In total 54 out of 67 sources have this information available, while for the
remaining 13, all not FIR detected, 6 have an SED fit in the analysis of Suh et al. (2017), 
but no significant torus component, and 7 do not have the minimum of 
five photometric band detections required in Suh et al. (2017) to perform the SED fitting.

Fig.~\ref{fig:lumlum} shows the distribution of the AGN Mid-IR luminosity, vs. 
the observed (left) and absorption-corrected (right) \lum\ for the sample of CT AGN (filled circles), compared 
with the 1048 sources in the COSMOS Legacy catalog with \liragn\ available (gray dots).
The magenta curve shows the relation between intrinsic \lum\ and \liragn\ published in Stern (2015).
The dispersion around this relation for the M16 sample is $\sigma=0.38$.

It is generally assumed that CT AGN have the same distribution as the M16 sample. However, as 
can be seen from the right panel, 90\% of the CT AGN selected here lie above the 
average \lum-\liragn\ relation, with an average offset of 0.7dex, 
and there are seven sources that lie at about $3\sigma$ from the relation.
While for some of these sources it is possible that the correction for absorption is over-estimated
(notice also the large uncertainties on the intrinsic luminosities), Fig.~\ref{fig:lumlum} suggests that CT 
sources with intrinsic \lum\ below the Stern (2015) relation
are missed because their observed \lum\ would cause them to be below the selection threshold applyed here based on number of 
counts, or even below the detection threshold for the survey.
Indeed there are 40 soures with 5-30 net counts and \liragn\ available (not shown here), 
that lie below the curve for log\nh=24 in
the left panel, and that would be missing in the right panel, since we do not have a reliable estimate of the intrinsic 
\lum.

As a further proof of this, we used the properties of 7 sources lying at $\sim1\sigma$ above 
the relation for the intrinsic \lum (from Fig.~\ref{fig:lumlum} right panel), 
to simulate what would be the observed count rate, 
if such sources were at $1\sigma$ below this relation (i.e. 0.8 dex lower intrinsic \lum). 
These sources cover a range of z and \nh\ representative of the entire sample. 
The typical number of counts from the simulated spectra 
(all the parameters being the same with the exception of the intrinsic \lum)
is $\sim10$ net counts, with SNR $\sim1.5$ for 6 out of 7 sources, 
and in 1 case the simulated spectrum is below the background level. 
Therefore CT source $1\sigma$ below the Stern et al. relation would be excluded from our 
analysis due to a low number of counts, or even undetected in the Cosmos Legacy catalog itself.

This suggests that a sizable fraction of CT AGN with similar properties
but lower intrinsic luminosity, with respect to the detected ones, is still missing due to their 
low X-ray fluxes. This selection effect, due to the flux limit of the survey, 
will be taken into account (see Appendix C) when computing intrinsic CT fractions.

%%%%%%%%%%%%%%%%%%%%%%%%%%%%%%%%%%%%%%%%%%%%%%%%%%%%%%%%%%%%%%%%%%%%
\begin{figure}
\begin{center}
\includegraphics[width=7cm,height=6.5cm]{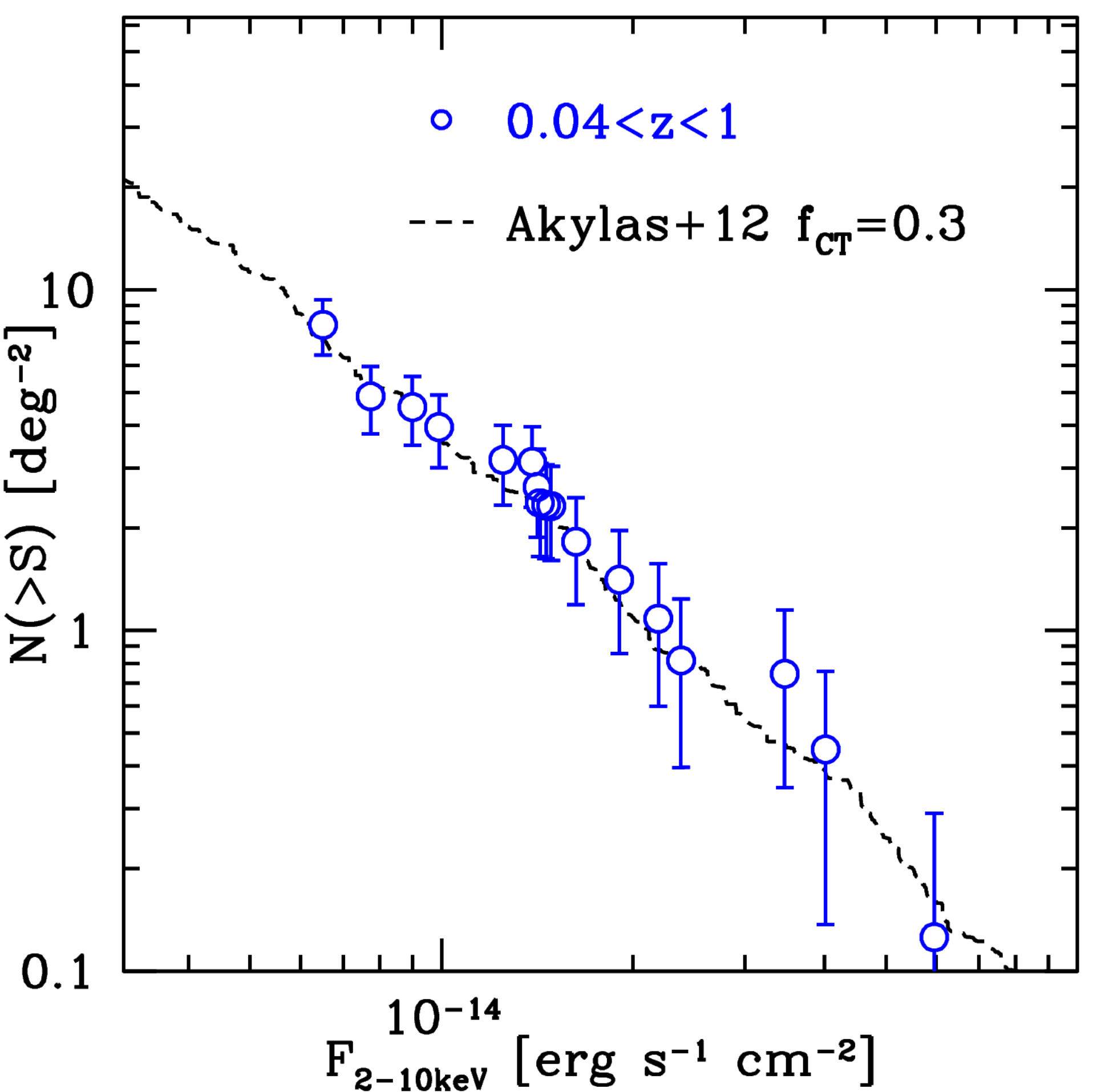}\vspace{0.2cm}
\includegraphics[width=7cm,height=6.5cm]{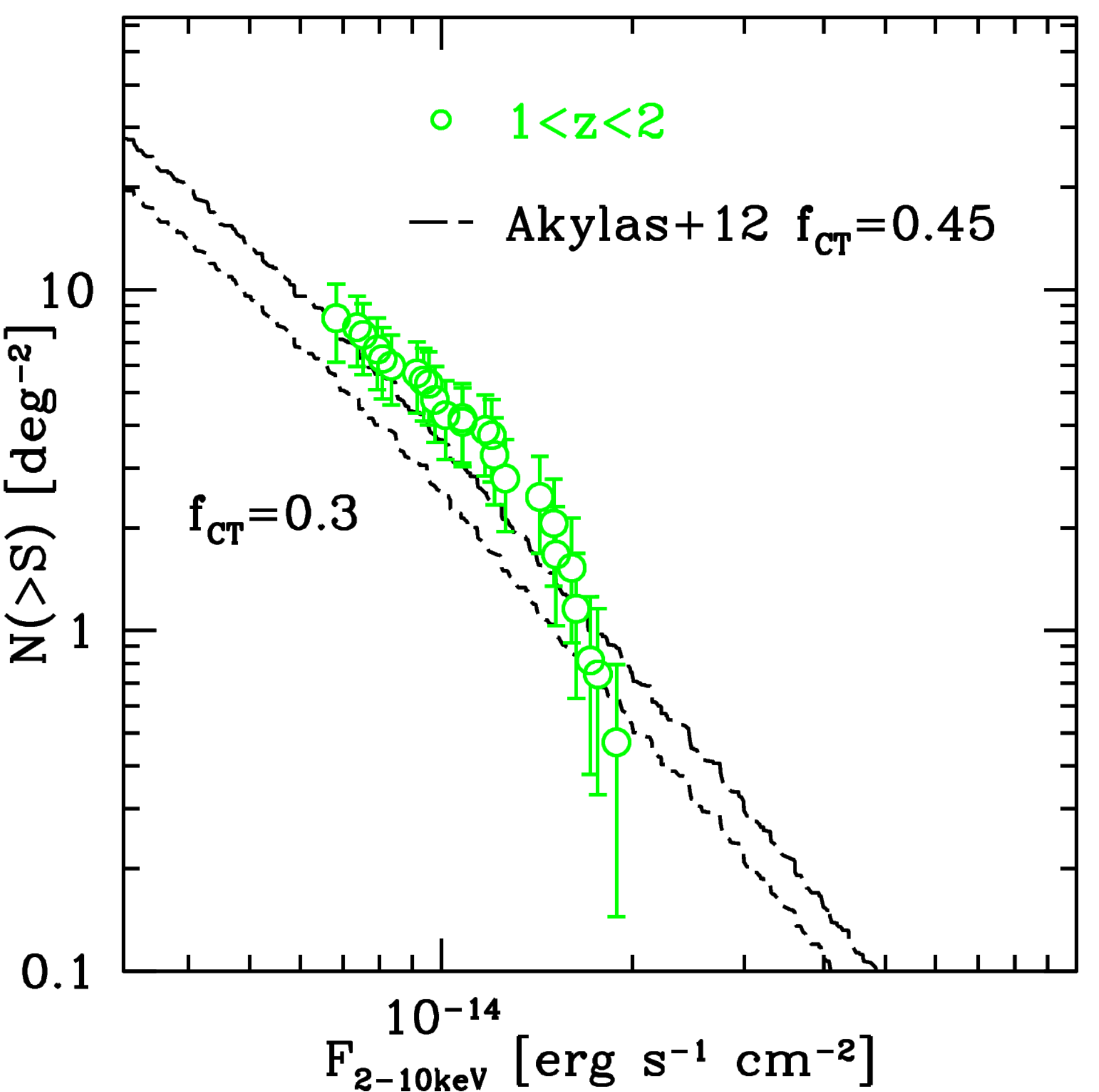}\vspace{0.2cm}
\includegraphics[width=7cm,height=6.5cm]{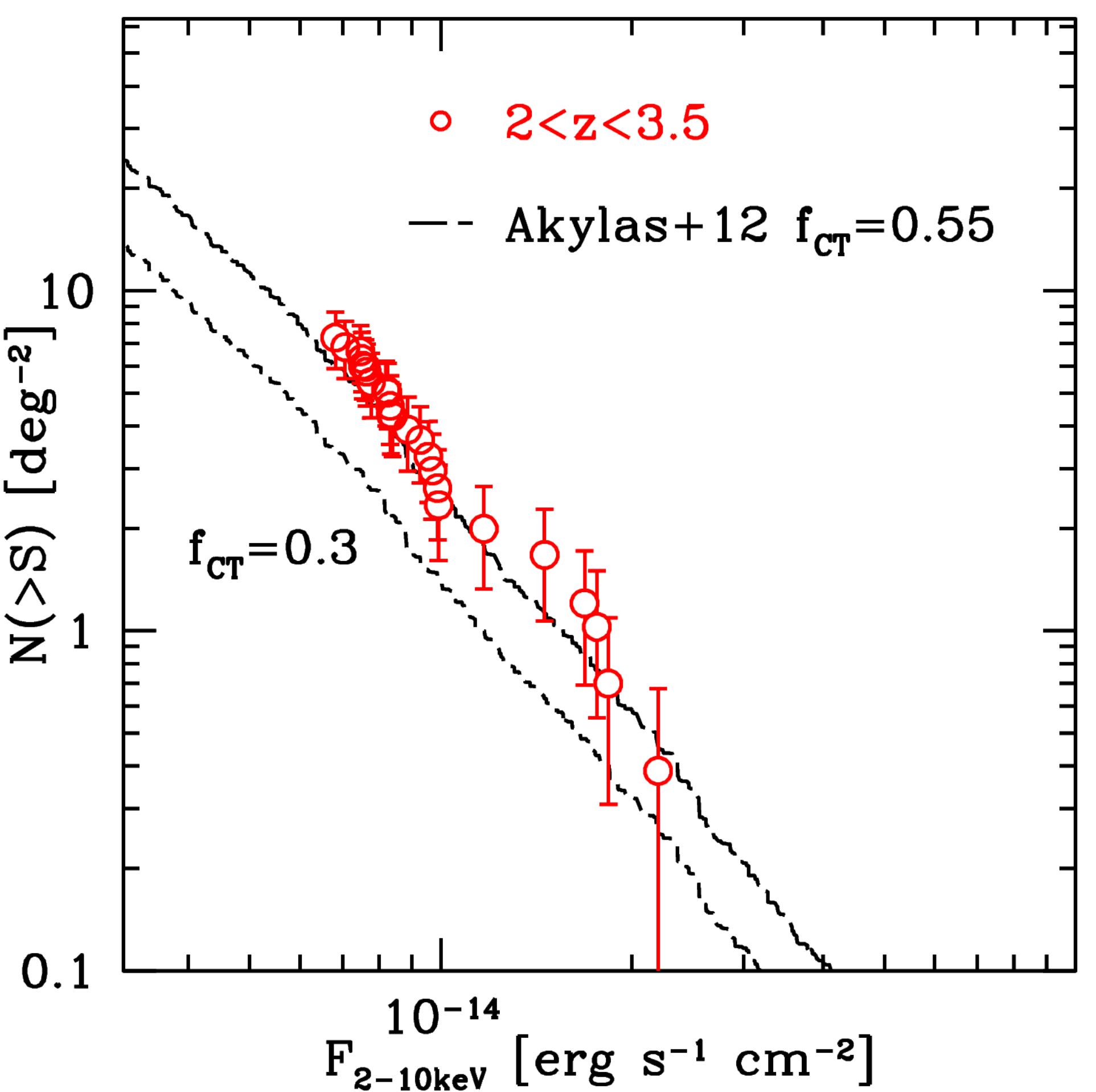}
\caption{{log}N-{log}S of CT candidates at $0.04<z<1$ (top), $1<z<2$ (center) and $2<z<3.5$ (bottom), corrected 
for survey sensitivity and classification bias, compared with the model of Akylas et al. 2012. 
The short dashed curves show the 
expected {log}N-{log}S for a constant \fct=0.3, while the long dashed curves in the center and bottom panel
show the increased \fct\ needed to qualitatively match the observed data points.}
\label{fig:lognlogs}
\end{center}
\end{figure}
%%%%%%%%%%%%%%%%%%%%%%%%%%%%%%%%%%%%%%%%%%%%%%%%%%%%%%%%%%%%%%%%%%%%%%%

%%%%%%%%%%%%%%%%%%%%%%%%%%%%%%%%%%%%%%%%%%%%%%%%%%%%%%%%%%%%%%%
\begin{figure}
\includegraphics[width=8.5cm]{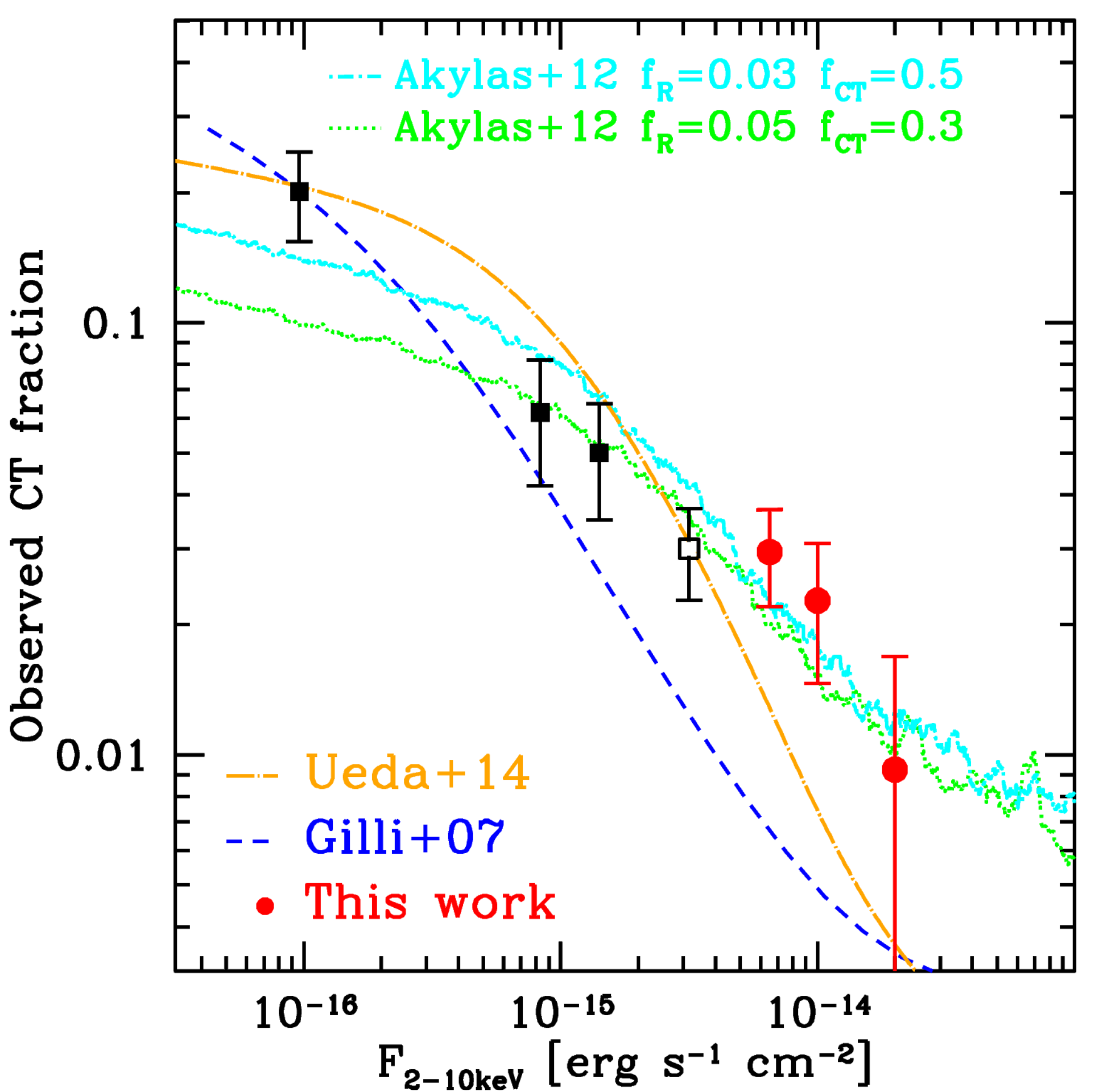}
   \caption{Observed \fct\ as a function of observed flux for the COSMOS Legacy CT candidates (red circles) 
   compared with data points from the literature, and with different CXB predictions.
  The black squares report the observed \fct\ from BU12 (CDFS 4Ms data), Brunner et al. (2008, Lockman 
  Hole), Tozzi et al. (2006, CDFS 1Ms data) and Hasinger et al. (2007, XMM-COSMOS, through hardness ratio, empty square) 
  from low to high flux.
}
\label{fig:ctf_flux}
\end{figure}
%%%%%%%%%%%%%%%%%%%%%%%%%%%%%%%%%%%%%%%%%%%%%%%%%%%%%%%%%%%%

We can also derive a selection efficiency in the region below the cyan dotted line in the left panel of Fig.~4 
(i.e. observed \lum\ below what expected for log\nh=24 \cm2),
which is often adopted in the literature for selection of CT candidates.

While in the local Universe it is possible to sample \lum-\liragn\ offsets of up to 2-3 dex 
(Gandhi et al. 2009, Asmus et al. 2015) making this selection effective (see also La Caria et al. 2018 submitted),
in high-z surveys the depth of the X-ray observation does not 
allow to sample such high \lum-\liragn\ offsets, translating into a limited efficiency
(see e.g. Lanzuisi et al. 2009, Georgantopoulos et al. 2011) or large fractions of non-detection
(Alexander et al. 2008, Del Moro et al. 2016, see Goulding et al. 2011 for a low-z example on shallow X-ray data).

Indeed we can show that the efficiency of CT selection for our sample is only $\sim33\%$ (5.3 effective CT candidates among 16 
total sources below the selection threshold), while it misses 90\% of the CT candidates selected on the basis of the spectral analysis. 
Therefore, the selection of CT AGN based only on the distance from the intrinsic relation, 
in high-z samples in medium-deep X-ray surveys, is not efficient because of the large disperion in the intrinsic 
relation and because CT sources lying below the intrinsic relation tend to be not detected in the X-ray catalog.

\section{Number density of CT AGN}

Taking advantage of the large sample of CT AGN selected through the analysis described in Sec.~3,
we derived number counts for our CT AGN sample in three redshift bins: $z=0.04-1$, $z=1-2$, and $z=2-3.5$
(hereafter z1, z2 and z3 respectively).
Each source is counted only for the fraction of PDF above $10^{24}$ \cm2. 
We apply the corrections described in Appendix C1 and C2 (differential sky coverage and classification bias, respectively),
computed as a function of source redshift and flux.

The {log}N-{log}S in the three bins is shown in Fig.~\ref{fig:lognlogs}. 
The error in each data point is computed taking into account only the Poissonian 
error related to the number of sources observed.
We compare our data points with the model predictions from Akylas et al. (2012). 
In this model it is possible to modify the intrinsic fraction of CT \fct, 
together with other parameters such as power-law photon index,
high energy cut-off (\ecut) and reflection fraction ($f_R$), 
expressed as the ratio between reflection and intrinsic continuum fluxes in the 2-10 keV rest frame band\footnote{see the on-line tool at \url{http://indra.astro.noa.gr/xrb.html}}.
%%%%%%%%%%%%%%%%%%%%%%%%%%%%%%%%%%%%%%%%%%%%%%%%%%%%%%%%%%%%

\begin{figure}
\includegraphics[width=9.cm]{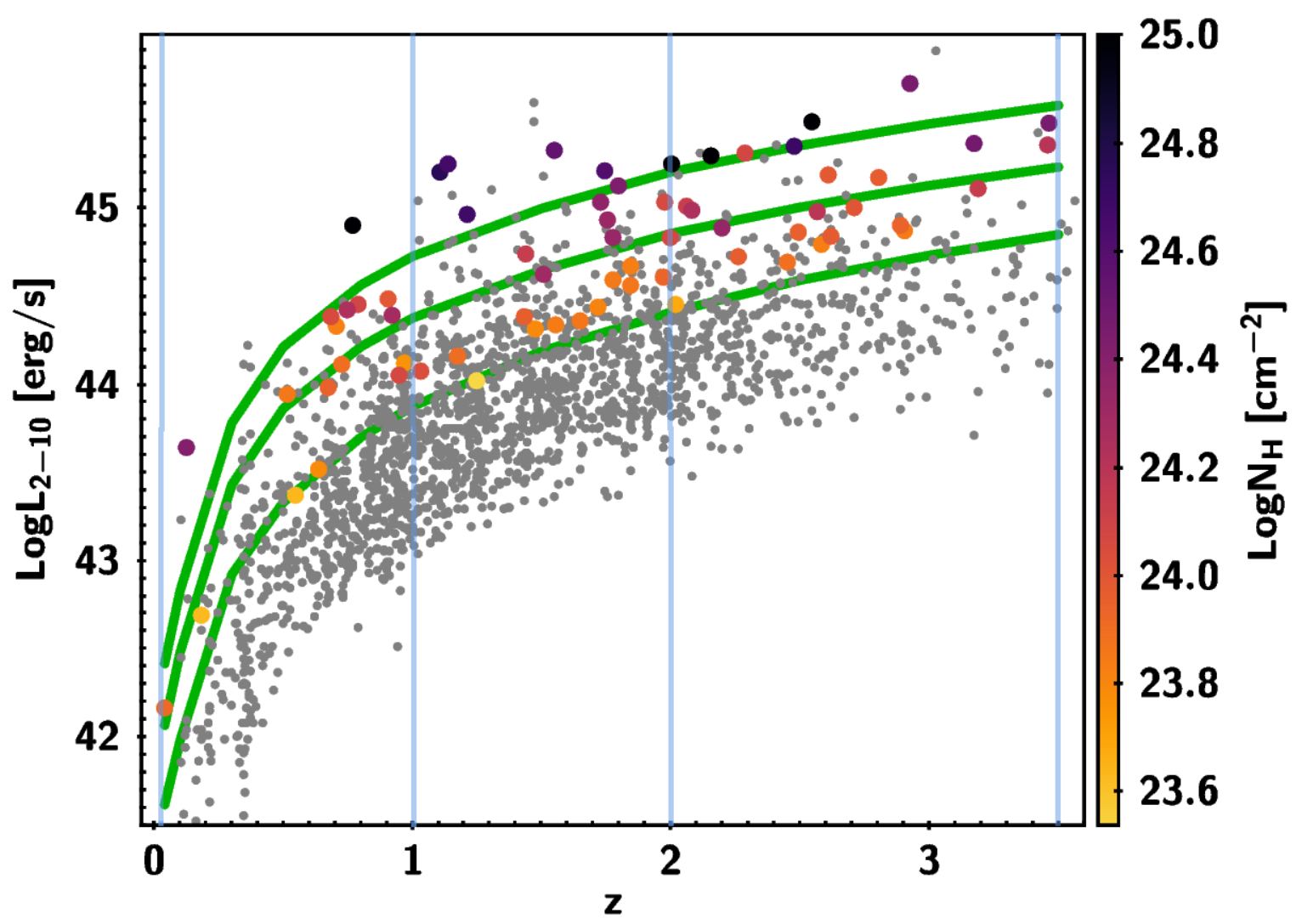}
    \caption{Distribution of intrinsic 2-10 keV luminosity vs. redshift for the CT AGN candidates (color coded for {log}(\nh)),
    and for the sample in M16 (gray dots). The vertical blue lines mark the limits of the 3 redshift bins.
    The green curves show the limit of intrinsic luminosity we are sensitive to, for {log}(\nh)=24,24.5,25 (\cm2) given
    the survey flux limit.}
    \label{fig:z_lum}
\end{figure}
%%%%%%%%%%%%%%%%%%%%%%%%%%%%%%%%%%%%%%%%%%%%%%%%%%%%%%%%%%%%

From this model, we derived the {log}N-{log}S for sources in the \nh\ bin 24-26 assuming 
$\Gamma=1.8$, \ecut$=195$ keV, $f_R=0.03$
(consistent with the reflection fraction of $\sim3\%$ derived from the model described in Sec.~2.2), 
in the three redshift bins. 
Keeping all the other parameters constant, \fct\ must increase from 0.3 in z1, 
to 0.45 and 0.55 in z2 and z3 respectively,
to match the observed distribution at those redshifts. All these values of \fct\ 
are still within the $1\sigma$ c.l. contours for the poorly constrained fraction of CT AGN derived from 
the Akylas et al. (2012) fit for the XRB intensity as a function of energy.
A more quantitative estimate of the evolution of \fct\ with redshift will be derived in Sec.~5 
using a different approach.

Using the {log}N-{log}S from the full COSMOS Legacy catalog (Civano et al. 2016) 
we can derive the observed fraction of CT AGN as a function of observed 2-10 keV flux.
This quantity is useful since it allows for comparisons with both data points from the literature and
CXB models.
Fig.~\ref{fig:ctf_flux} shows the observed \fct\ in three flux bins for the COSMOS Legacy sample (red circles). 
The orange curve shows the prediction from the base-line CXB model of Ueda et al. (2014), that derives an intrinsic 
fraction N(24-26)/N(20-26) of $\sim0.33$.
The blue curve shows the G07 model, with an intrinsic N(24-26)/N(20-26) of $\sim0.37$ 
averaged over all luminosities.
The cyan and green curves show two different realizations of the Akylas et al. (2012) model. 
Both have $\Gamma=1.8$ and \ecut$=195$ keV, while $f_R$ goes from 0.03 (cyan) to 0.05 (green) and \fct\ 
from 0.5 (cyan) to 0.3 (green).
None of these models, however, allow for any evolution of the \fct\ with redshift.

%%%%%%%%%%%%%%%%%%%%%%%%%%%%%%%%%%%%%%%%%%%%%%%%%%%%%%%%%%%%%%%%%%%%
\begin{figure*}
\begin{center}
\includegraphics[width=7.5cm,height=7.5cm]{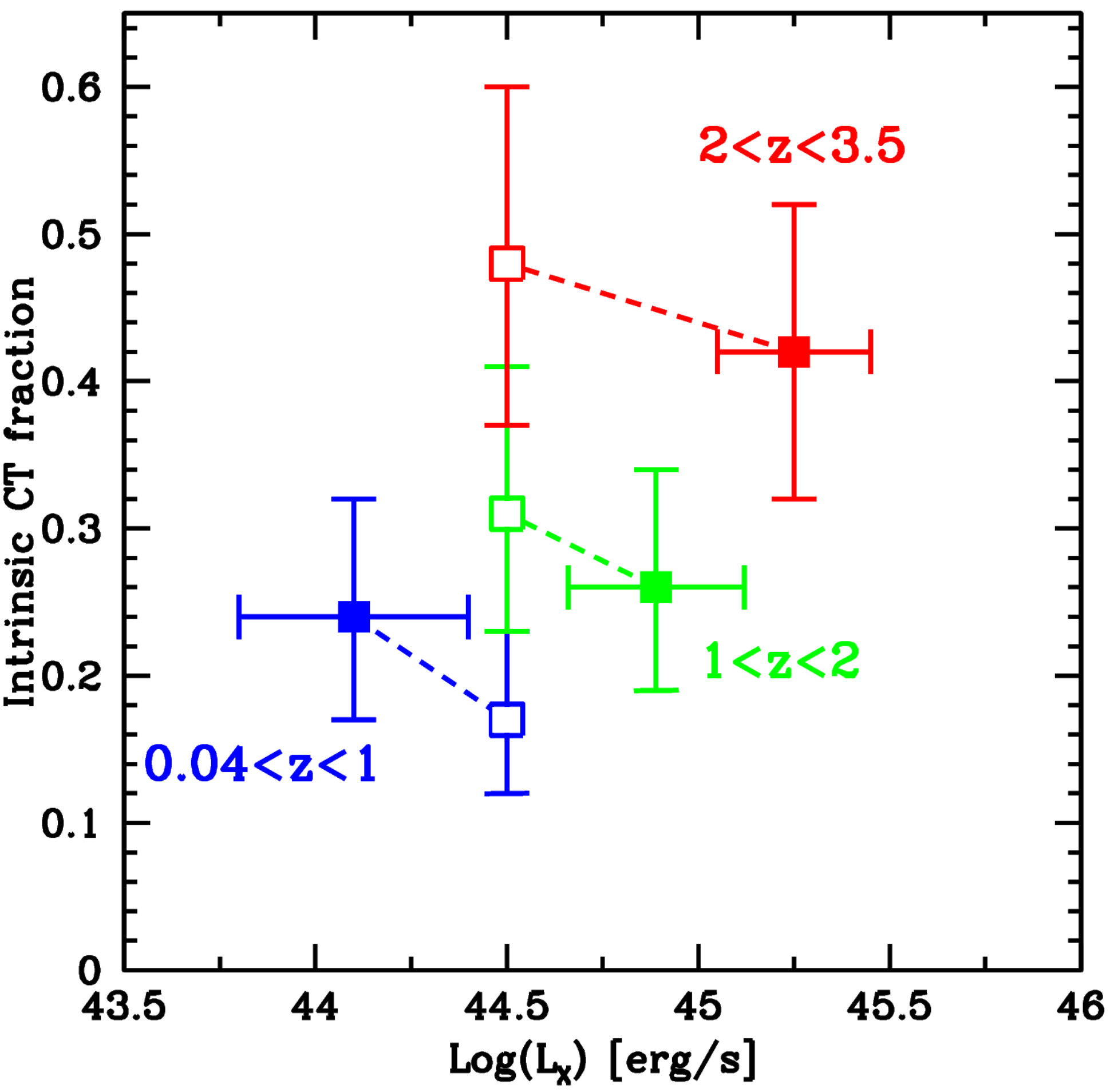}\hspace{0.5cm}
\includegraphics[width=7.5cm,height=7.5cm]{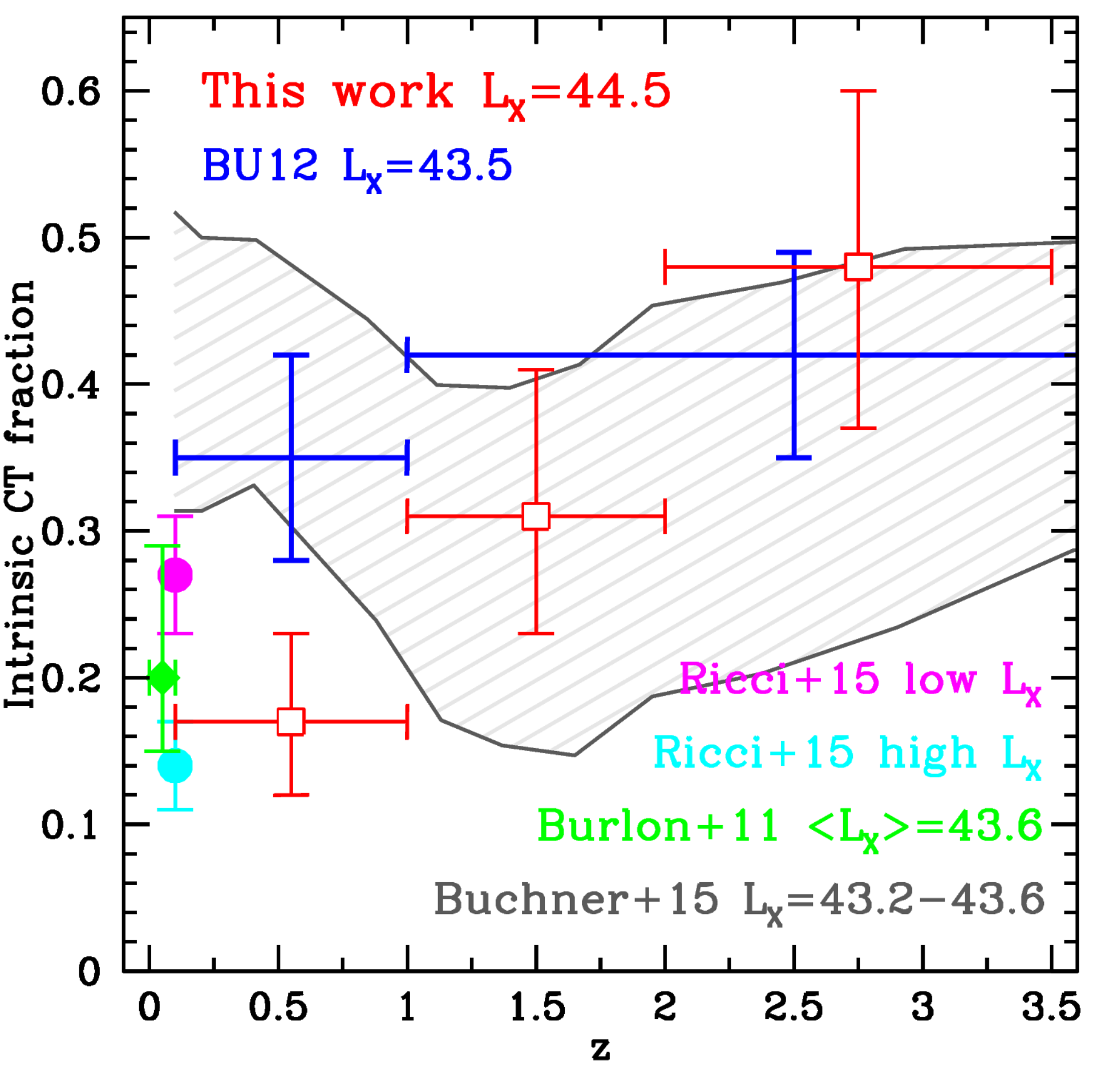}
\caption{{\it Left:}
Intrinsic fraction of CT AGN as a function of \lum, observed (filled squares) and extrapolated to {log}(\lum)=44.5 
erg/s  (empty squares), for the three redshift bins: blue for z1, green for z2 and red for z3.
{\it Right:} 
Intrinsic fraction of CT AGN as a function of redshift, 
for {log}(\lum)=44.5 erg/s (red empty squares). 
Data points from the literature are from Burlon et al. (2011 green diamond), 
Ricci et al. (2015 cyan and magenta circles) and BU12 (blue crosses).
The light gray dashed area is from Buchner et al. (2015).
}
\label{fig:ctf}
\end{center}
\end{figure*}
%%%%%%%%%%%%%%%%%%%%%%%%%%%%%%%%%%%%%%%%%%%%%%%%%%%%%%%%%%%%%%%%%%%%%%%

%%%%%%%%%%%%%%%%%%%%%%%%%%%%%%%%%%%%%%%%%%%%%%%%%%%%%%%%%%%%%%%%%%%%%%%
\begin{table*}
	\centering
	\caption{Intrinsic \fct\ at different luminosities for the three redshift bins.
	(1) Redshift range; (2) Number of CT AGN above the {log}(\nh)=24.5 (\cm2) completeness curve, 
	after correcting for classification bias; (3) Surface density of CT AGN, 
	given the survey flux-area curve; (4) Number of C-thin AGN above the {log}(\nh)=24.5 (\cm2) 
	completeness curve; (5)
        Surface density of C-thin AGN; (6) Average \lum\ and dispersion for the total sample;
        (7) \fct\ at the observed average {log}\lum; (8) \fct\ rescaled at {log}(\lum)=44.5 (erg/s). 
        The errors in this column also take into account the uncertainty on the slope F2 vs. \lum.}
	\label{tab:fct}
	\begin{tabular}{ccccccccc} % four columns, alignment for each
\hline
z             &  $N_{CT}$ &$N_{CT}$    & $N_{thin}$ & $N_{thin}$ &$<Log(L_X)>$          & \fct\   & \fct(44.5)   \\
              &           & deg$^{-2}$ &            & deg$^{-2}$ &                      &         &                   \\
 (1)          & (2)       & (3)        & (4)        &  (5)       & (6)                  & (7)     &   (8)          \\
\hline                            
$ 0.04 <z<1$  & 7.14        & 3.36        &  23        &   11.23    &  $44.1\pm0.3$     &   $0.24_{-0.07}^{+0.08}$      &    $0.19_{-0.06}^{+0.07}$      \\
$ \ 1<z<2 $   & 8.67        &4.29         &  25        &   11.80    &  $44.9\pm0.2$     &   $0.26_{-0.07}^{+0.08}$      &    $0.30_{-0.08}^{+0.10}$      \\
$ \ \ 2<z<3.5$& 9.51        &4.54         &  13        &    6.80    &  $45.2\pm0.2$     &   $0.42_{-0.09}^{+0.10}$      &    $0.48_{-0.11}^{+0.12}$    \\
\hline
\label{tab:tab2}
\end{tabular}
\end{table*}
%%%%%%%%%%%%%%%%%%%%%%%%%%%%%%%%%%%%%%%%%%%%%%%%%%%%%%%%%%%%%%%%%%%%%%%

Note that the increase from $f_R=0.03$ (cyan) to 0.05 (green)
compensates for the decrease in \fct, especially at the bright fluxes, 
and therefore both realizations of the Akylas et al. model are in agreement with our data points. 
The cyan curve (lower $f_R$ and higher \fct) is in better agreement with 
the lowest flux data point coming from CDFS-4Ms (BU12), while the green curve (higher  
$f_R$ and lower \fct) is in better agreement with other data points from the literature at higher fluxes.
We note that the flux ratio between direct and reflected components in the 2-10 keV band,
derived from the {\it Mytorus} model, with the parameters described in Sec.~2, 
and log(\nh)$=24$ (\cm2) is 0.033, close to the values for the cyan curve.

The discrepancy of the observed \fct\ predicted by different models at high fluxes is mainly due to the different assumptions 
on the reflection component: in G07 the reflection contribution is different 
for obscured and unobscured sources,
and is neglected for luminous sorces. In Akylas et al. (2012) $f_R$ is the same for all sources.
We also add that in both cases, the reflection component is modelled with a disk-reflection model (i.e. {\it PEXRAV} in {\it XSPEC}),
while for CT sources the reflection produced by a toroidal structure would be more appropriate  and produce
a different spectral shape (see MY09).

\section{Intrinsic Fraction of CT AGN}

We derive the intrinsic fraction of CT AGN in three redshift bins, in a common \lum\ range, 
following the procedure described in BU12. 
In order to compute intrinsic fractions of CT sources in z and \lum\ bins, we have to build samples that are 
complete with respect to a given value of \nh.
Therefore we compute the sensitivity curves shown in Fig.~\ref{fig:z_lum},
by converting the flux limit of our CT sample, {log}F$_{2-10 keV}>-14.2$ (\cgs)
into an intrinsic luminosity limit, for a given redshift, computed adopting a spectral model
as described in Sec.~2.2, and with {log}(\nh)$=24$, $24.5$ and $25$ (\cm2), respectively.

All the sources from the full COSMOS Legacy catalog above each of the three curves
shown in Fig.~\ref{fig:z_lum} constitute a complete sample up to that \nh. 
Unfortunately, the curve for {log}(\nh)$=25$ (\cm2) 
includes very few sources, and therefore the determination of the intrinsic \fct\ would have very large
uncertainties. We therefore decided to use the complete sample above the curve for {log}(\nh)$=24.5$ (\cm2),
and computed the fraction of CT AGN using the samples (in each redshift bin) above the curve for {log}(\nh)$=24.5$ \cm2. 
This represents a lower limit of the full intrinsic \fct\ defined as N({log}\nh$=24-25)/N(tot)$, since we can still miss some sources
with {log}(\nh)$>24.5$ \cm2 close to the completeness curve.
The fractions that are obtained for the sample above {log}(\nh)$=25$ (\cm2) are however very similar, although with large error bars,
in all the three z bins. Therefore we argue that the \fct\ derived for the sample complete up to {log}(\nh)$=24.5$ (\cm2)
is a good approximation for the full CT sample.

Tab.~\ref{tab:tab2} summarizes the results obtained following this approach. 
The number of CT AGN is corrected accounting for the classification bias and survey sensitivity.
The average log \lum\ and its dispersion (at $1\sigma$ c.l.) is derived from the full sample (CT+C-thin) in each redshift bin.
Due to the small sample sizes, we adopted the Bayesian approach presented in Cameron et al. (2011)
to derive confidence intervals (at $1\sigma$ c.l.) on the observed ratio \fct.

To rescale \fct\ to a common \lum\ range, we exploit the well known 
linear relation between the fraction of obscured AGN \f2, defined as \f2 $= N(22-24)/N(20-24)$,
and {log}(\lum) (see e.g. Hasinger et al. 2008, BN12 and  Ueda et al. 2014).
As done in BN12 we adopted the slope $0.281\pm0.016$ found in Hasinger (2008) on a sample of $>1000$ AGN up to $z=5$. 
In the following we take into account the slope uncertainties in the extrapolation to a common \lum\ range.

We derive \fct\ (i.e. $N(24-26)/N(20-26)$) at a given \lum\ using the relation
observed between \f2 and \lum, assuming a flat \nh\ distribution above 
{log}(\nh)$=22$ \cm2 (i.e. $N(24-26) = N(22-24)$, and therefore \fct=\f2/(1+\f2).
This assumption is supported by the fact that the \nh\ distribution, observed in the CT sample and control sample 
together, is indeed nearly flat in the bins {log}\nh$(22-24)$-{log}\nh$(24-26)$ 
(we observe a ratio of 1.2-1.0-0.9 in the three redshift bins).
For each redshift bin, we therefore use the average \lum\ of the CT+C-thin sample, and extrapolate  
\fct\ to {log}(\lum) $=44.5$ erg/s in each bin using these relations.

The results of this approach are shown in Fig.~\ref{fig:ctf} (left) for the three redshift bins.
The filled squares represent the measured \fct\ for each redshift bin, sampling increasing luminosities 
at increasing redshift. The empty squares are the values of \fct\ at {log}(\lum) $=44.5$ erg/s estimated with the method
explained above.

%%%%%%%%%%%%%%%%%%%%%%%%%%%%%%%%%%%%%%%%%%%%%%%%%%%%%%%%%%%%%%%
\begin{figure*}
\includegraphics[width=17.5cm,height=4.5cm]{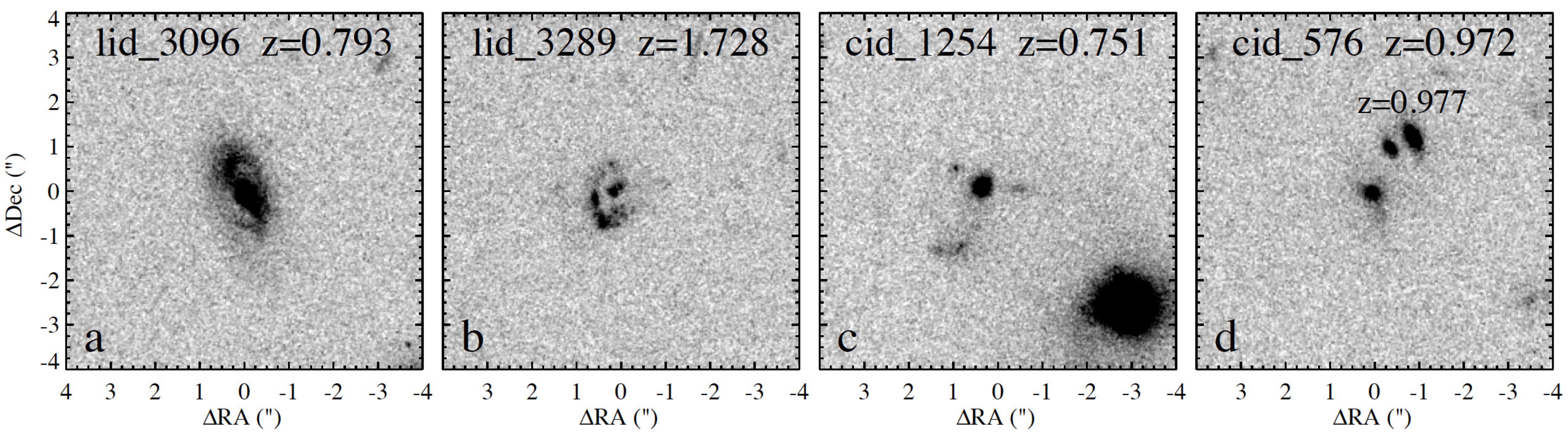}
   \caption{HST/ACS I-band cutouts (Koekemoer et al. 2007) of four CT AGN candidates. 
   They show examples of an isolated spiral (barred) galaxy (a) a merging system (b) a galaxy with tidal tails 
   (c) and a group of close companions with similar redshifts (d). The cut-outs have a size of $8\times8$ arcsec.
}
\label{fig:cutouts}
\end{figure*}
%%%%%%%%%%%%%%%%%%%%%%%%%%%%%%%%%%%%%%%%%%%%%%%%%%%%%%%%%%%%55

%%%%%%%%%%%%%%%%%%%%%%%%%%%%%%%%%%%%%%%%%%%%%%%%%%%%%%%%
\begin{figure}
\begin{center}
\includegraphics[width=\columnwidth]{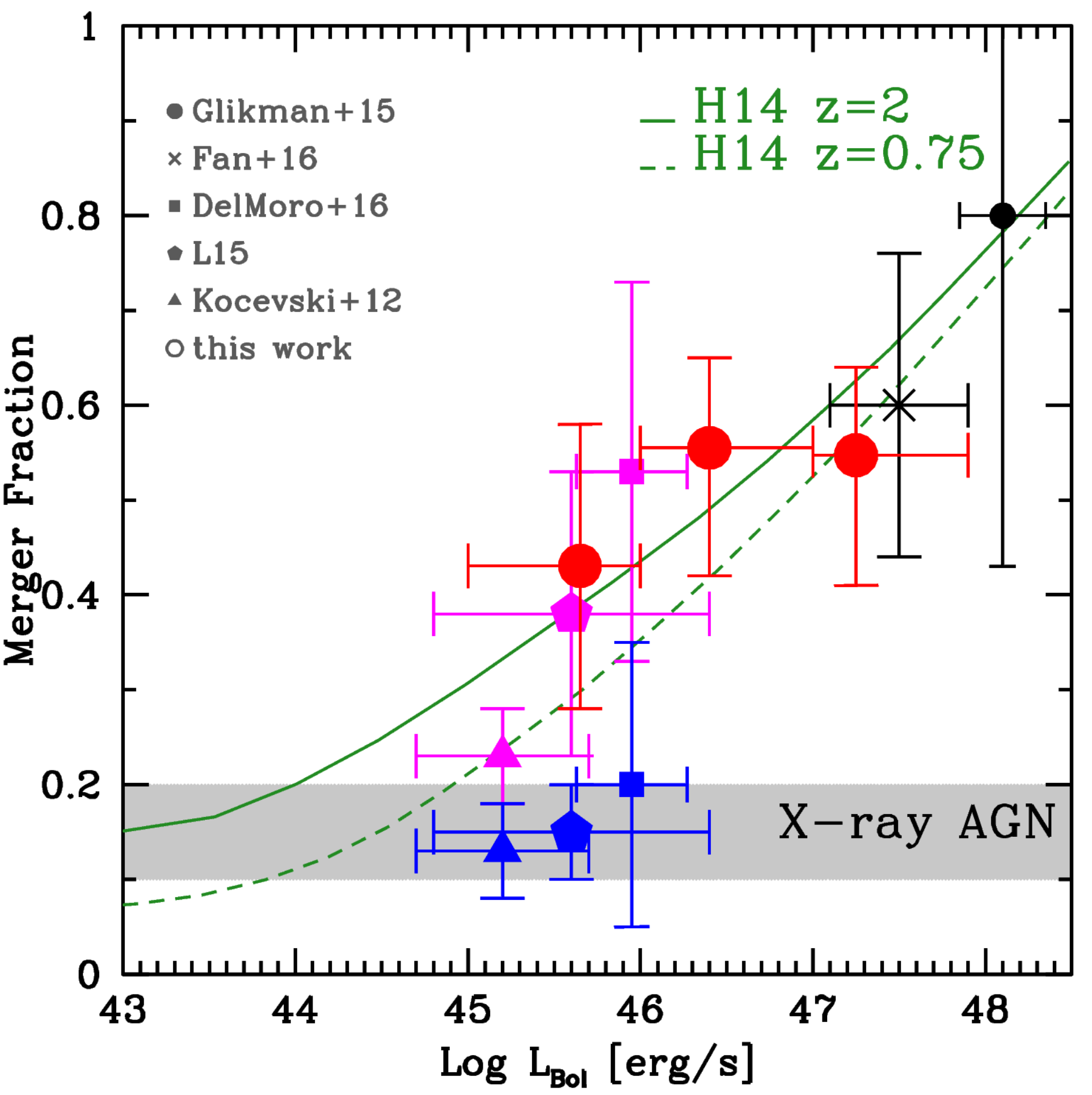}
\caption{Fraction of merger/disturbed morphology for several samples of CT and non-CT AGN at different \lbol.
The red circles show the measurements for our sample, divided in three \lbol\ bins.
The magenta points show fractions for different CT AGN samples in the literature.
Blue points are the parent sample of non-CT AGN in each of these studies.
The black points at high luminosities are merger fractions for 
dust-reddened quasars (Fan et al. 2016, Glikman et al. 2015) at $z\simgt2$.
The green thick (dashed) line shows the prediction of the Hickox et al. (2014) model of AGN-galaxy 
coevolution, at $z=2$ ($z=0.75$).
}
\label{fig:merg}
\end{center}
\end{figure}
%%%%%%%%%%%%%%%%%%%%%%%%%%%%%%%%%%%%%%%%%%%%%%%%%%%%%%%%%%%%%%%%%%%%%%%

We note that, given the definitions above, \f2\ cannot be $>1$ (the extreme case in which all
the Compton thin sources are obscured above $10^{22}$ \cm2) and therefore \fct\ cannot be $>0.5$, by construction.
In this regime the model is clearly an oversimplification; for example,
it is possible that the relation \f2-{log}(\lum) evolves with redshift, 
or that the assumption $N(24-26) = N(22-24)$
is no longer valid.

% Applying instead the shape of the relation from G07 plotted in Fig.~\ref{fig:ctf} (left) to the observed data points, would
% tranlates into slightly lower values of \fct\ at {log}(\lum)=43.5 erg/s for all the redshift bins 
% (\fct=0.31, 0.35 and 0.51 for z1, z2 and z3, respectively), but the global picture would not change significantly.

Fig.~\ref{fig:ctf} (right) shows the evolution of \fct\ with redshift, as derived in the left panel, 
for {log}(\lum)=44.5 erg/s for the COSMOS Legacy sample (red empty squares).
We compare our results with others from the literature: in particular, the blue error-bars show the results
from BU12 from the 4Ms CDFS data set at {log}(\lum)=43.5 (erg/s); magenta and cyan circles show the 
results from Ricci et al. (2015) from the hard-band selected, \swift-BAT sample in the local Universe\footnote{
We recall, however, that Marchesi et al. (2018) showed that these low-z results may be  
overestimated by $\sim20\%$ based on \nus\ data. But see also La Caria et al. 2018 submitted for a smaller discrepancy in the other direction between pre- and post-\nus\ data.}  
, at low ({log}$L_{14-195}=40-43.7$ erg/s) and high ({log}$L_{14-195}=43.7-46$ erg/s) luminosities, respectively;
the green diamond is the measurement from Burlon et al. (2011) on an earlier version
of the \swift-BAT hard band catalog, with no luminosity cut. The light gray shaded area is from
Buchner et al. (2015) for {log}\lum$=43.2-43.6$ erg/s.
All these measurements point toward an increase of \fct\ from low to high redshift (with the exception of the Buchner et al.
results).

 Performing a linear fit between \fct\ and z to our data points at high luminosity, 
we get a slope of $m\sim0.13$. For an easier comparison with literature results we compute the dependency adopting 
an expression such as \fct$=\beta(1+z)^\alpha$.
The best fit parameters are  $\beta=0.11_{-0.04}^{+0.05}$ and $\alpha=1.11_{-0.12}^{+0.14}$. Interestingly, this relation
perfectly fits also the data point at high luminosity from Ricci et al. (2015, cyan square in Fig.~\ref{fig:ctf} right). 
The slope of \fct\ as a function of redshift at high luminosities is therefore significantly steeper than the slope found in
BU12 with similar methods for lower luminosites ($\alpha\sim0.3$), suggesting a faster evolution of \fct\ for more luminous systems.
For comparison the evolution of the obscured fraction (again at low luminosites) measured in Liu et al. (2017), is fitted with best fit parameters
$\beta=0.42\pm0.09$ and $\alpha=0.60\pm0.17$.

\section{Morphology of CT AGN}

We collected all the available HST ACS I band images from the COSMOS mosaic (Koekemoer et al. 2007, I band magnitude limit 
27.2) for the 67 CT AGN candidates. 
We adopted a simplified visual classification scheme comprising, on one side, isolated/undisturbed morphologies,
and, on the other side, sources clearly in merger state or with post-merger features such as tidal tails 
and disturbed morphologies or with close companions\footnote{In five cases the CT AGN has a companion within 3''
that is bright enough to have a photometric redshift in the COSMOS2015 catalog (Laigle et al. 2016),
and we checked that in all cases the two sources have comparable redshifts.}.
17 sources have no classification since the HST counterpart is absent 
or too faint to determine any morphological feature ($I>25$), 16 out of these 17 have $z>1.5$.

Our morphological classification is supported by the comparison with the Tasca Morphology Catalog
(Tasca et al. 2009). All the 50 CT AGN with visual classification have a counterpart in the catalog,
and 90\% of the sources classified by us 
as merging or post-merging systems (panels b and c of Fig.~\ref{fig:cutouts} respectively)
are classified as irregulars in at least two of the three classification schemes of that catalog, 
based on different parametric estimates.

\lbol\ for all the sources has been derived using the Marconi et al. (2004) \xray\ bolometric correction.
The final sample of 50 sources with morphological information has been divided in \lbol\ bins of one dex width.
The bin  {log}(\lbol)=44-45 erg/s comprises only three sources and is therefore ignored.
As in the previous sections, each of the CT AGN in our sample has been weighted by the fraction of its \nh\ 
PDF above $10^{24}$ \cm2, 
and this number has been corrected for classification bias and survey sensitivity.
The errors are computed adopting the Bayesian approach of Cameron et al. (2011). 

We compared our results in Fig.~\ref{fig:merg} with results from
several previous studies, and from the Hickox et al. (2014) model.
The literature results at intermediate luminosities ({log}\lbol$=45-46$ erg/s, Kocevski et al. 2012, L15 and Del Moro et al. 2016) 
give a merger fraction both for the CT candidate samples (magenta) and for the C-thin parent samples (blue) at the same luminosity,
while for very high luminosities (black points) morphological results refer to 
$z\simgt2$ dust-reddened quasars (Fan et al. 2016, Glikman et al. 2015), not necessarily CT.
The results from our large sample of CT AGN (red points) allow to confirm that highly obscured AGN at high 
luminosities show an increase in merger fraction with respect to the C-thin parent sample.
Our results are also in agreement with the AGN-galaxy coevolution model of Hickox et al. (2014),
in which galaxy mergers play a prominent role in triggering the most luminous and obscured AGNs at $z\sim2$.
We note that the average redshift of the bins {log}(\lbol)$=45-46$, $46-47$ and $47-48$ erg/s are $z=0.93$, $1.83$ and $2.38$ respectively.
Also the data points from the literature at the highest luminosities (from Fan et al. 2016 and Glikman et al. 2015) are derived for sources
at $z>2$. Therefore with these data sets it is not possible to disentangle a luminosity from a redshift dependence of the merger fraction.

\section{Conclusions}

We compiled one of the largest sample of CT AGN candidates at high redshift (67 individual sources), 
from the COSMOS Legacy point source catalog. For comparison, there are 100 CT AGN candidates in 
Brightman et al. (2014) summing three \chandra\ surveys and 165 CT AGN candidates in Buchner et al. 2015 from four different surveys.
Our sample was selected applying a physically motivated model for the \xray\ emission, 
and MCMC methods to efficiently explore the parameter space \nh\ vs. \lum\ and taking into account the full PDF of these
quantities in our analysis. 
The number of 'effective' CT AGN derived from the sum of the \nh\ PDF above $10^{24}$ \cm2\ is 38.5, 41.9 after correction
for identification bias.
The two main results of this work are:

\begin{itemize}
\item The fraction of CT AGN increases as a function of redshift, 
from $\sim0.2$ at $z=0.1-1$, to $\sim0.3$ at $z=1-2$ to $\sim0.5$ at $z=2-3.5$.
These values are derived from the observed fraction of CT AGN 
in each redshift bin, rescaled to a common luminosity range of log(\lum)=44.5 erg/s. 
This evolution can be parametrized as \fct$=0.11(1+z)^{1.11}$.
A similar trend, with values of 0.30, 0.45 and 0.55 was found for the global CT fraction by a qualitative 
comparison of the logN-logS of the CT AGN in the same redshift bins, with the CXB model of Akylas et al. (2012).\\

\item The fraction of CT AGN in merging/interacting systems is systematically higher, by a factor 2.5-3, than
that observed in the parent sample of C-thin AGN in several other studies. 
This increase also has a positive dependence on the \lbol.
Given the redshift and luminosity distribution of the samples, it is not possible to disentangle a luminosity from a redshift dependence of the merger fraction.
\end{itemize}

Our interpretation of these results, together with other pieces of evidences in the literature
(see Alexander \& Hickox 2012 and Netzer 2015 for recent reviews) is that at high redshift,
the simplest version of the ``unification scheme'' for AGN (Antonucci 1993) does not hold anymore,
and the orientation with respect to a nuclear obscuring torus is not the
main driver of the differences between obscured and unobscured AGN.
Instead, the conditions of the AGN host galaxy, i.e. amount and distribution of cold gas, nuclear star formation,
level of interaction with neighbors etc., may play a major role in determining the
amount of obscuration we measure in the \xray.

Such ``large scale'' obscuration should exist in addition to
the classical nuclear dusty region commonly referred to as torus 
(see e.g. Elvis 2012). The relative importance of these different components will depend 
on the host galaxy environment, gas and dust properties, 
that at high redshift may favour the presence of large amount of obscuring material 
(e.g. an augmented gas fraction, see e.g. Carilli \& Walter 2013, Scoville et al. 2017, Darvish et al. 2018). 

Some recent results from AGN-host population studies also point in this direction, 
deriving a positive correlation between obscuration, at least in the Compton-thin regime,
and host gas mass or stellar mass (Rodighiero et al. 2015, Lanzuisi et al. 2017).
As an extreme example for the CT regime, Gilli et al. (2014) showed that the CT absorber 
observed in an obscured QSO hosted in an ultraluminous infrared galaxy at $z=4.75$,
can be fully accounted for by the amount of gas and its compactness measured
by ALMA.

Indeed the gas fraction increases from $z=0$ to $z=3$ derived from ALMA dust continuum observations 
(Scoville et al. 2014, 2017, Darvish et al. 2018) goes in this direction.
We envisage that direct cold gas mass and size measurement through ALMA CO observations in CT AGN hosts 
at high redshift will help to shed light on the nature of the nuclear absorber and possibly its connection with
the host star formation properties (e.g. Perna et al. 2018 submitted).

While the \chandra\ deep fields such as CDFS, CDFN and COSMOS represent the current limit for the search of CT AGN
at high redshift, large samples (hundreds) of such sources will be
detected and routinely identified (Georgakakis et al. 2013) 
in the planned extragalactic survey (Aird et al. 2013) 
of the Wide Field Imager (WFI) on board the next generation ESA X-ray mission, Athena 
(Nandra et al. 2013).

As an example, the current {\it Athena}-WFI design will allow to collect, in the medium (600ks exposure) tier of the survey,
$\sim10^4$ full band net counts for a source like lid\_283. 
This is the highest redshift CT in our sample at $z=3.465$ and log(F$_{2-10})=-14$ \cgs.
Such spectral quality will result in a \nh\ uncertainty of $<5\%$.
Around $3500$ full band net counts will be collected for a source with similar intrinsic 
luminosity moved at $z=6$, with a derived \nh\ uncertainty of $\sim10\%$.

The NASA proposed {\it Lynx} mission (Gaskin et al. 2015) is planned to have a 0.5'' PSF, and the deep surveys 
perfomed with its High Definition X-ray Imager will be able to resolve high-z AGN where {\it Athena}
might be affected by source confusion, making possible to extend the constraints for the CT fraction at 
even fainter luminosities and higher redshifts.

\section*{Acknowledgements}

The authors thank the referee for the constructive comments and recommendations
which helped to improve the readability and quality of the paper.
This research has made use of
data obtained from the Chandra Data Archive and software
provided by the Chandra X-ray Center (CXC) in the CIAO
application package.
The author acknowledges financial support from the CIG grant ``eEASY'' n. 321913, 
from ASI-INAF grant n. 2014-045-R.0 and from PRIN-INAF-2014 (``Windy Black Holes combing galaxy evolution'').

%%%%%%%%%%%%%%%%%%%%%%%%%%%%%%%%%%%%%%%%%%%%%%%%%%



%\bibliographystyle{mnras}


%%%%%%%%%%%%%%%%% APPENDICES %%%%%%%%%%%%%%%%%%%%%

\appendix

\section{Parameter space exploration}

We adopted a Markov Chain Monte Carlo approach to explore the parameter space in 
the spectral fit.
Once the best fit with the standard C-stat likelihood is obtained, 
we run the  MCMC code implemented in {\it XSPEC} (v. 12.9.1).
We use the Goodman-Weare algorithm (Goodman \& Weare 2010) 
with $10^4$ steps and $10^3$ burn in steps to ensure convergence and efficiently
explore the parameter space. The marginalization over the parameter of interest gives the full PDF distribution.
The errors reported in Tab.~1 are obtained by the classical command ``error'' in  {\it XSPEC}, 
but in this case the program uses the Chain to derive 
the confidence interval at a given c.l. directly from the PDF.

\begin{figure}
	\includegraphics[width=8.3cm,height=7cm]{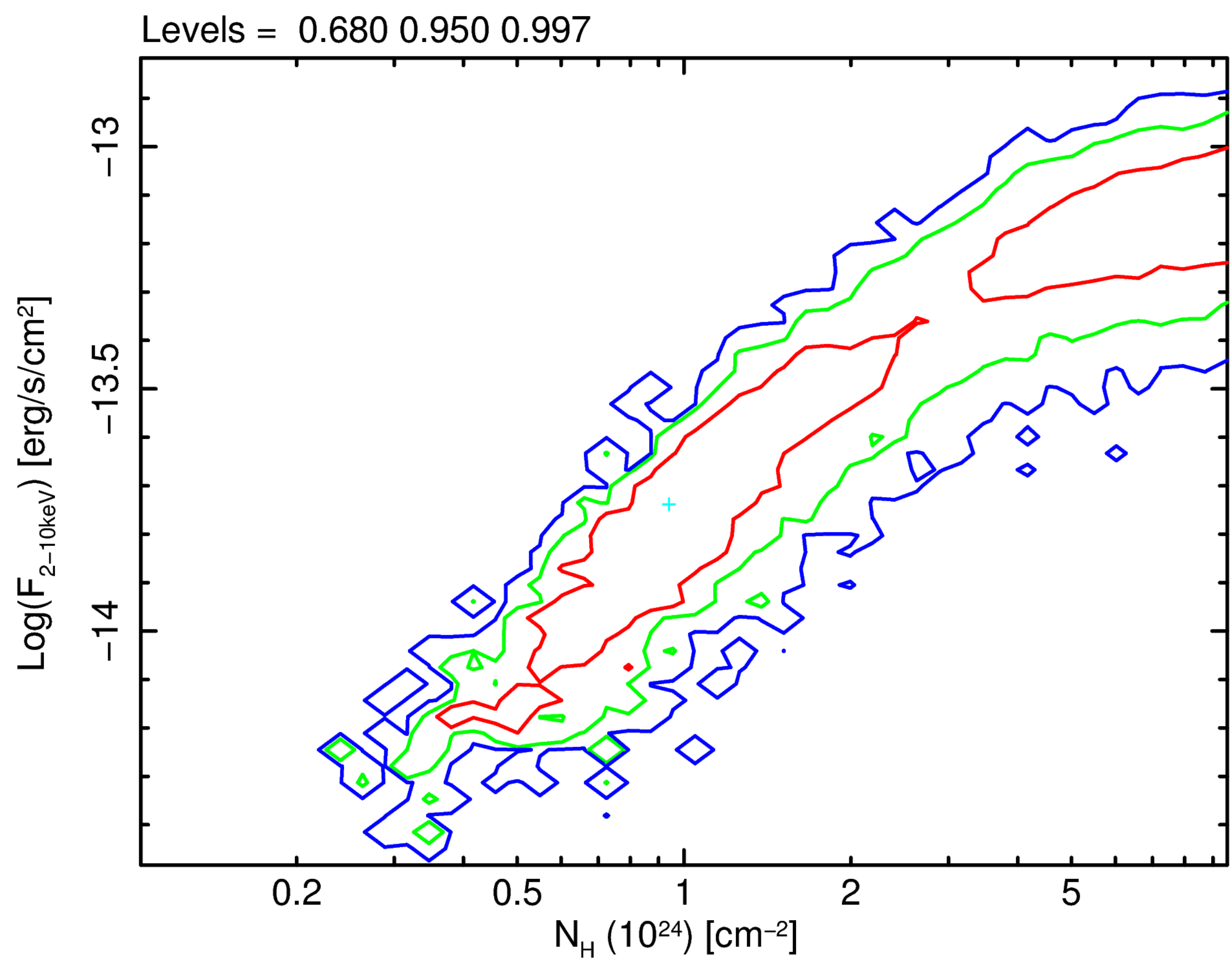}
    \caption{Probability distribution function for two parameters of interest, \nh\ and intrinsic Flux, for source lid\_3516.
    In this case there are two distinct minima of the fit statistic. The MCMC method is able to properly take into account both of them.}
    \label{fig:integ}
\end{figure}

Fig.~\ref{fig:integ} shows an example of a double-peaked PDF in the parameter space \nh\ vs. intrinsic flux, 
for source  lid\_3516. The red, green and blue contours show the 1, 2 and $3\sigma$ c.l., respectively.
The standard methods for error estimation would fail to correctly
estimate the uncertainties in these cases. 
Such a probability distribution in \nh\ is rather common in \xray\ spectra of 
CT candidates, that often allow for two solutions, one with \nh\ below the CT threshold 
and lower \lum\, and a second at higher \nh\ and intrinsic luminosity (see also Buchner et al. 2014).

Finally, we computed the fraction of PDF that each 
source shows above the CT threshold, and 
then apply a probabilistic approach when analyzing the number counts 
of samples of CT AGN: the {\it effective} number of CT is the sum of the fraction of the PDF above CT values
for each source in the sample.

%%%%%%%%%%%%%%%%%%%%%%%%%%%%%%%%%%%%%%%%%%%%%%%%%%%%%%%%%%%%%%%%%%%%
%%%%%%%%%%%%%%%%%%%%%%%%%%%%%%%%%%%%%%%%%%%%%%%%%%%%%%%%%%%%%%%%%%%%

\section{Impact of different assumed parameters}

%%%%%%%%%%%%%%%%%%%%%%%%%%%%%%%%%%%%%%%%%%%%%%%%%%%%%%%%%%%%%%%%%%%%
\begin{figure*}
\begin{center}
\includegraphics[width=5.5cm,height=4cm]{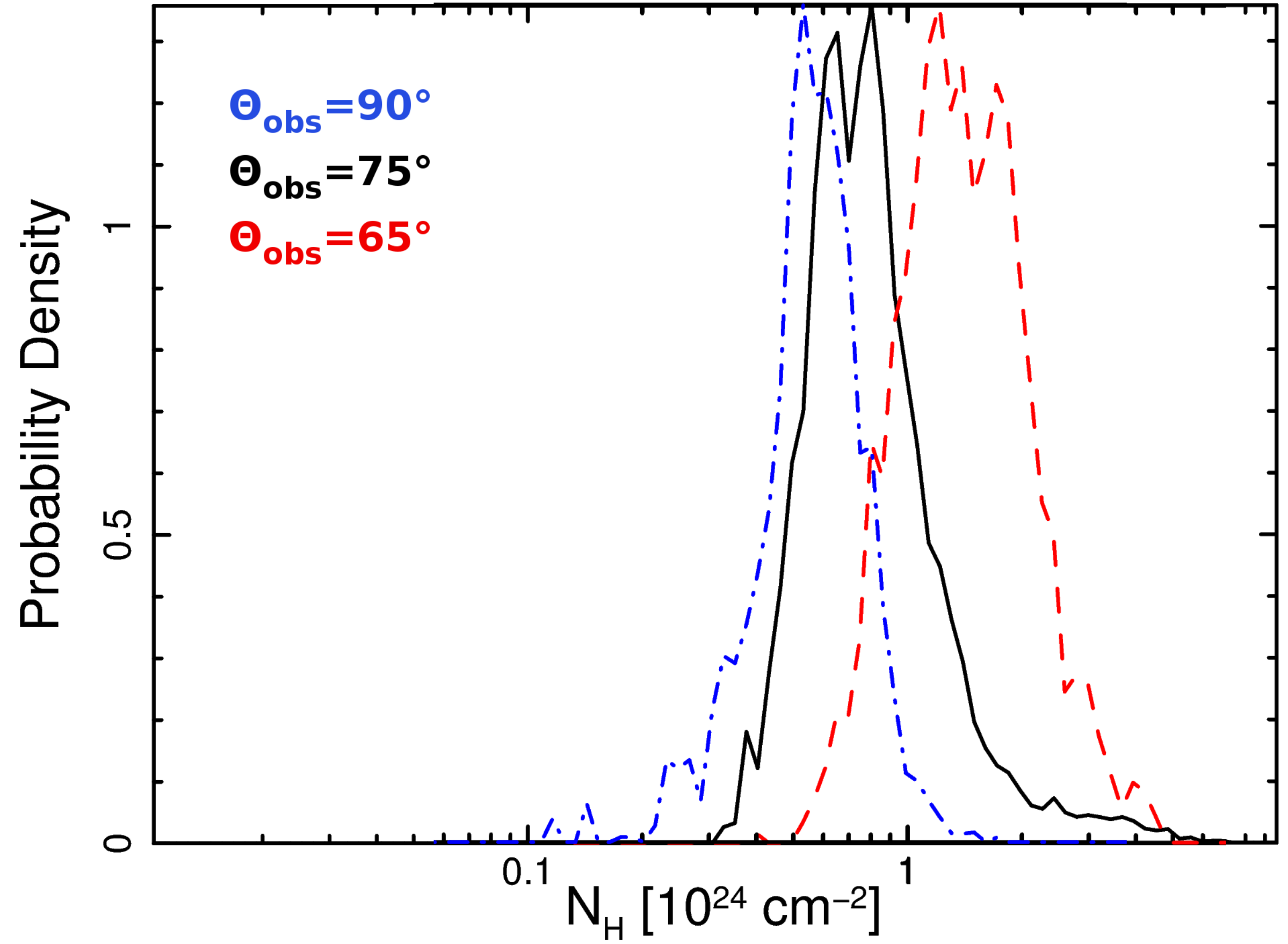}\hspace{0.2cm}\includegraphics[width=5.5cm,height=4cm]{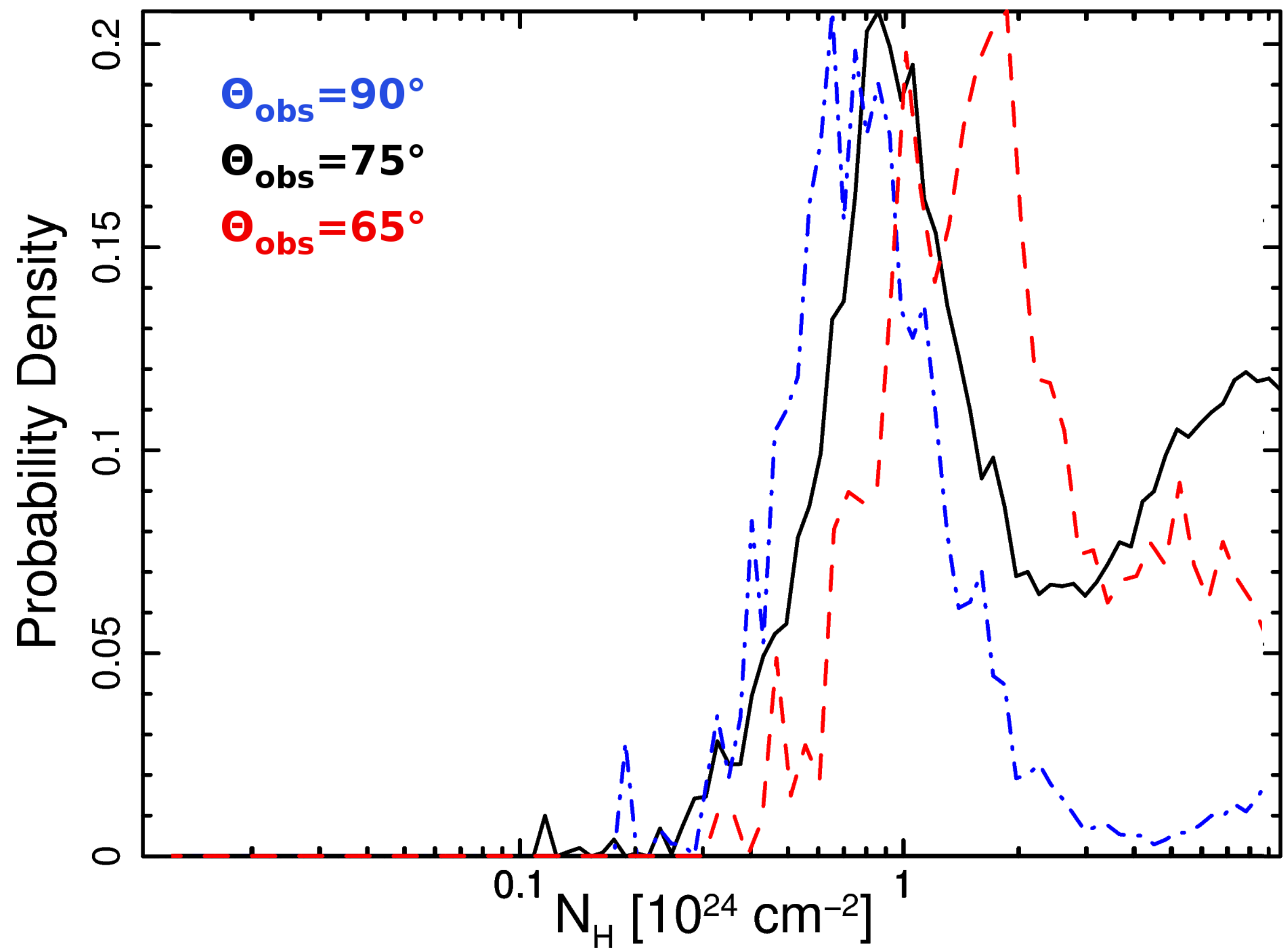}\hspace{0.2cm}\includegraphics[width=5.5cm,height=4cm]{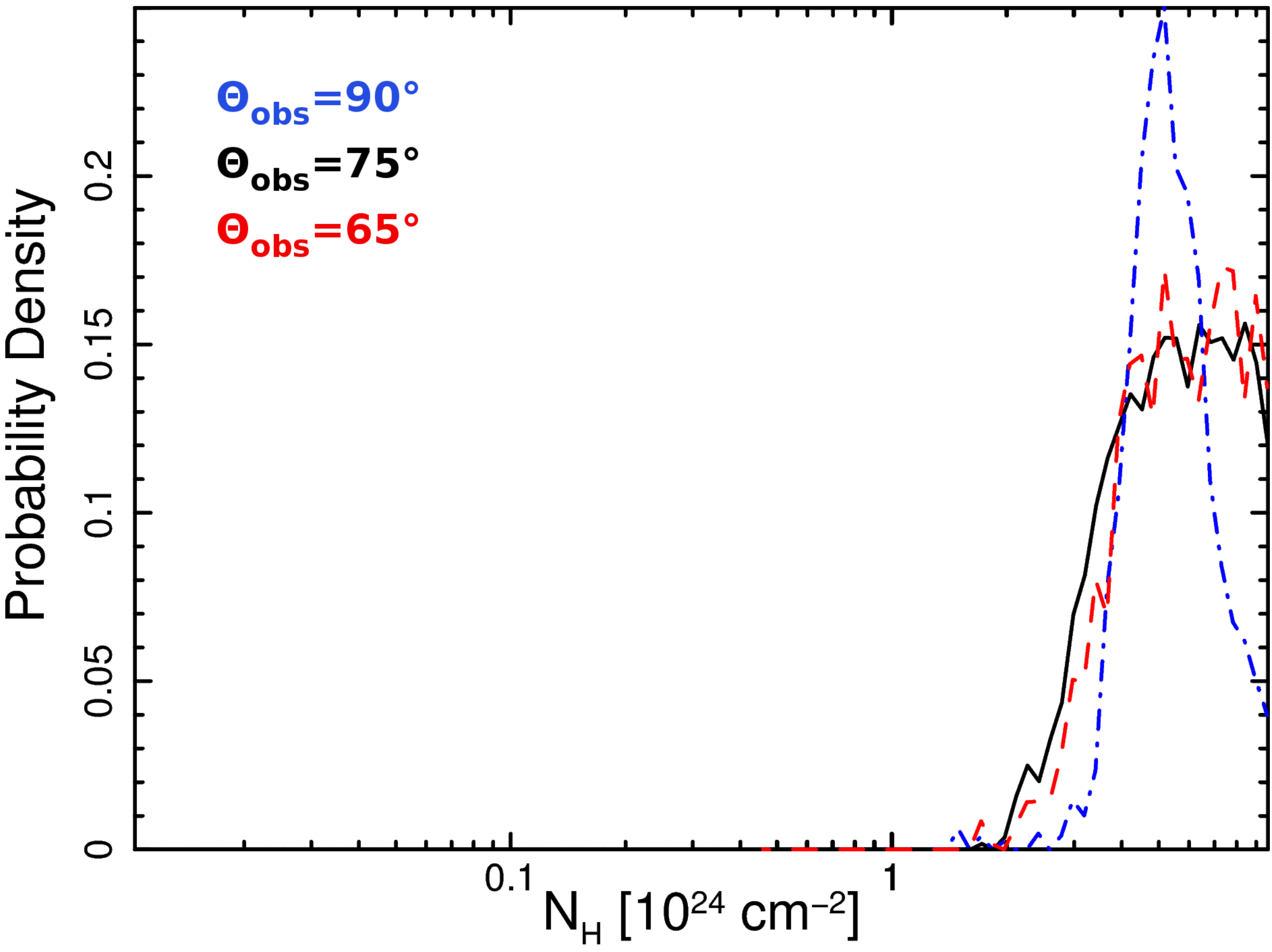}\vspace{0.3cm}
\includegraphics[width=5.5cm,height=4cm]{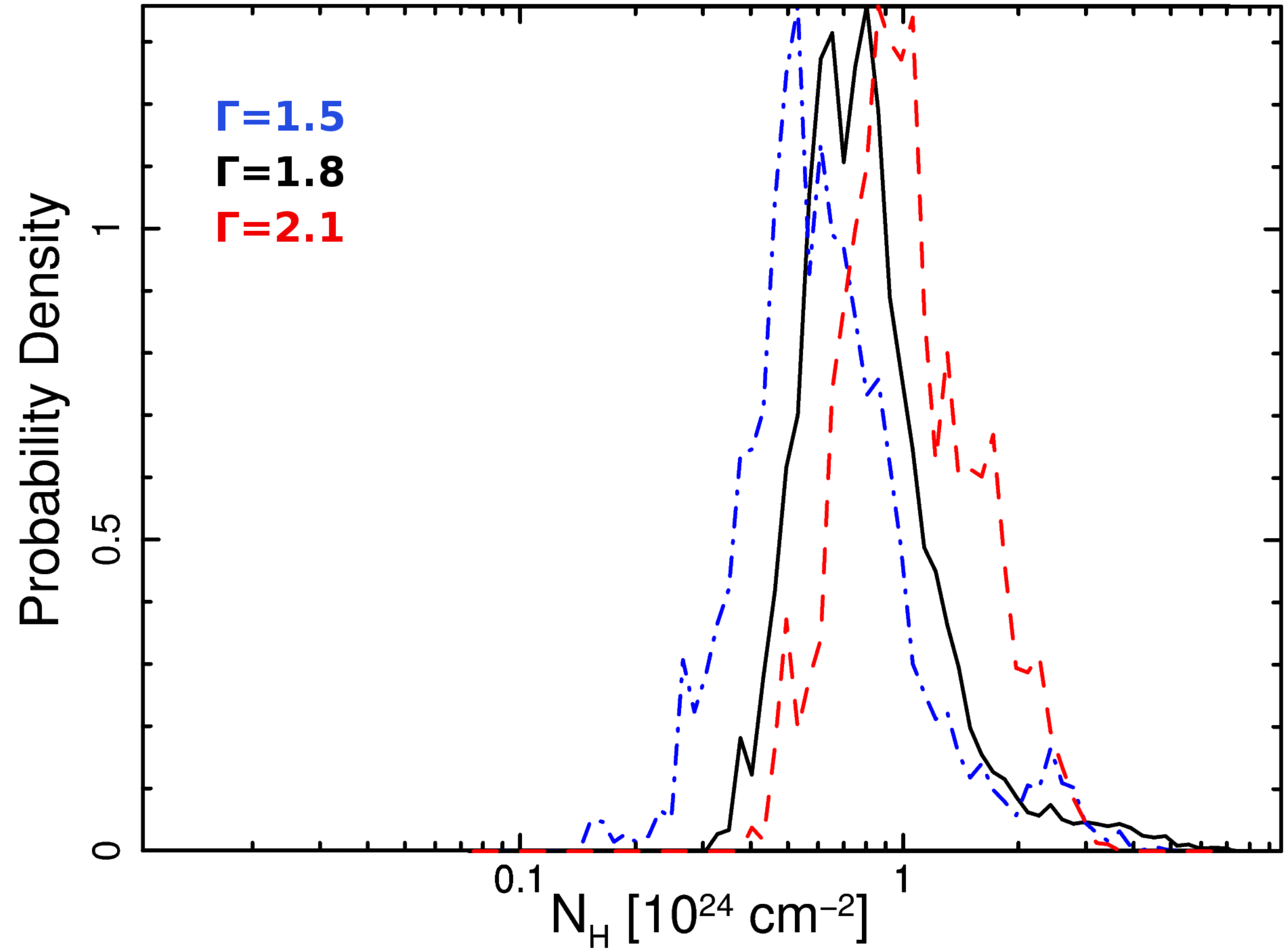}\hspace{0.2cm}\includegraphics[width=5.5cm,height=4cm]{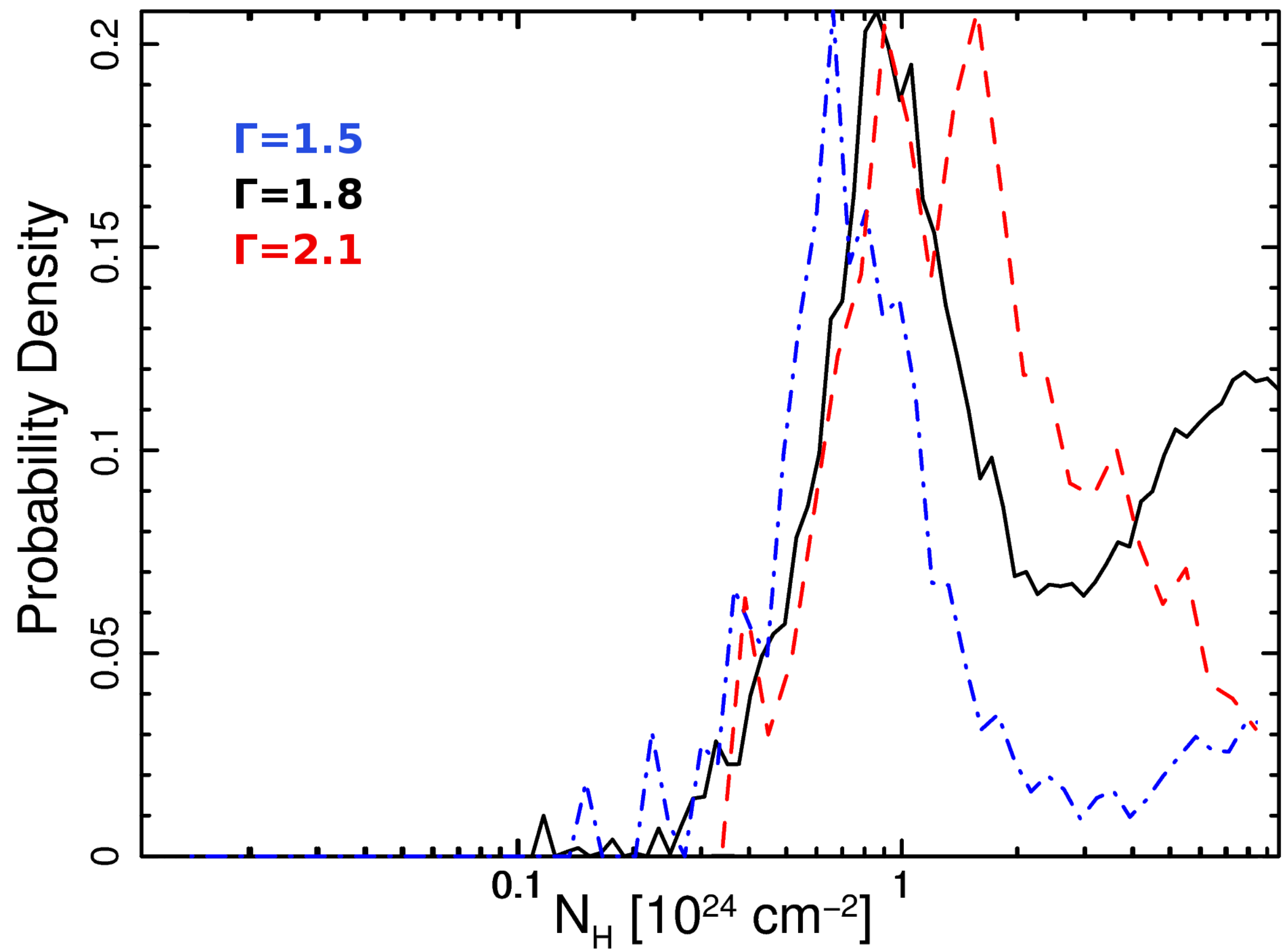}\hspace{0.2cm}\includegraphics[width=5.5cm,height=4cm]{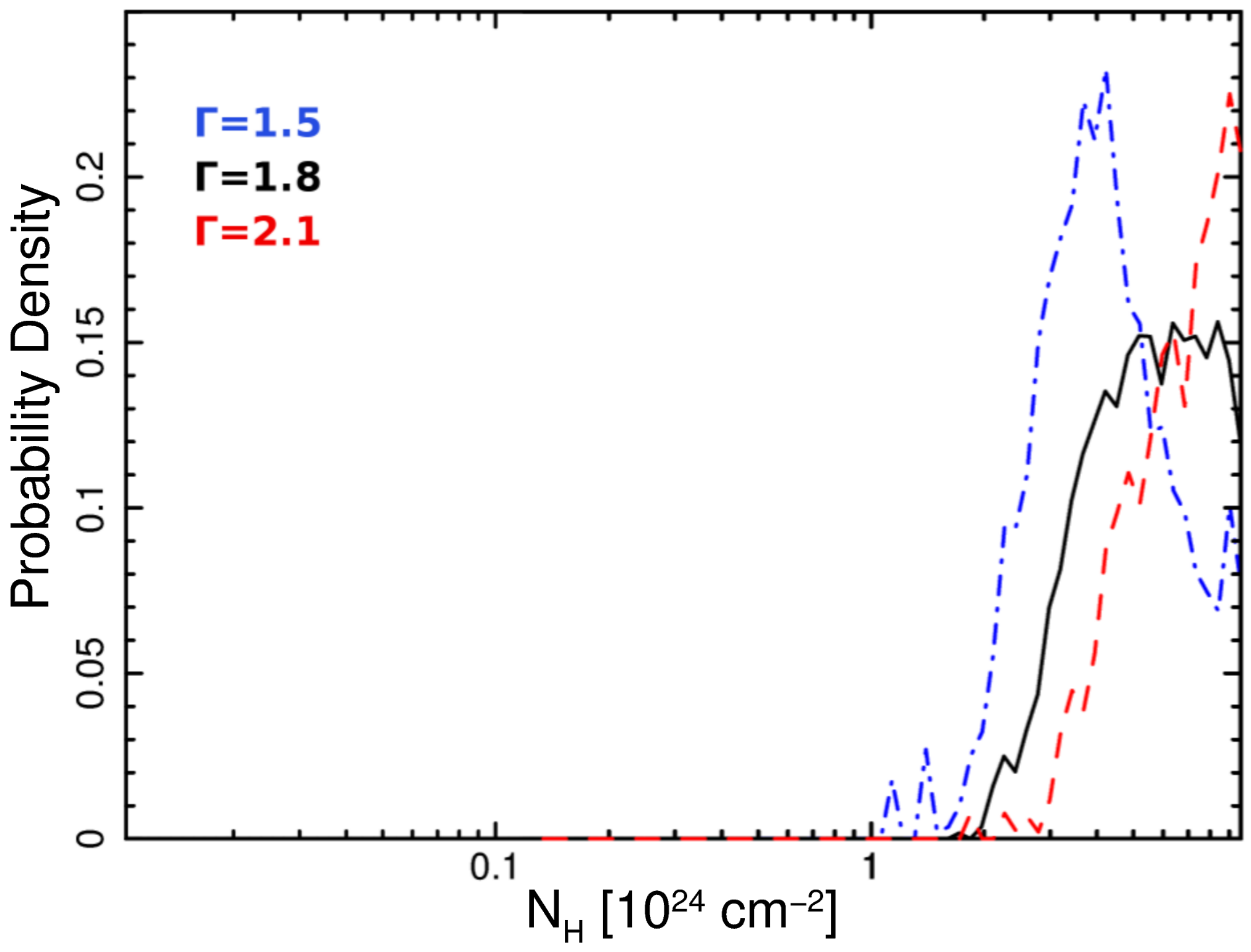}
\caption{{\it Top:} \nh\ PDF for different values of $\Theta_{obs}$, 65,75 and 90$^\circ$ (red, black and blue respectively),
for the sources shown in Fig.~2 (lid\_633, lid\_3516 and lid\_390). 
{\it Bottom:} \nh\ PDF for different values of $\Gamma$, 2.1, 1.8 and 1.5  (red, black and blue respectively).
}
\label{fig:app_pdfnh}
\end{center}
\end{figure*}
%%%%%%%%%%%%%%%%%%%%%%%%%%%%%%%%%%%%%%%%%%%%%%%%%%%%%%%%%%%%%%%%%%%%%%%

Here we show the impact of the use of different values of the fixed parameters in the model for CT AGN,
such as $\Gamma$ and inclination angle $\Theta_{obs}$, on the determination of \nh.
The top panels of Fig.~\ref{fig:app_pdfnh} show the variation of the \nh\ PDF for different assumed torus inclination angles: 
$\Theta_{obs}$=65,75, and $90^\circ$ (red, black and blue respectively) for the three sources shown in Fig.~2 
as representative of the sample.
The lower panels show the variation for different assumed photon indices: $\Gamma$=2.1, 1.8 and 1.5 (red black and
blue respectively) for the same sources.

The change in $\Gamma$ has the effect of shifting the \nh\ PDF by $\sim0.1$ dex in both directions.
The $\Gamma$ range explored corresponds to $\sim1.5\sigma$ of the observed distribution 
(e.g. Piconcelli et al. 2005, Bianchi et al. 2009).

The change in $\Theta_{obs}$ has a slightly larger effect, shifting the \nh\ PDF by $\sim0.1$ dex for $90^\circ$
and by $\sim0.3$ dex for $65^\circ$. This is due to the fact that, in the geometry of MYtorus model,
the LOS intercepts the torus only for $\Theta_{obs}>60^\circ$, and at $65^\circ$ the section of the torus 
intercepted is very small. Therefore the best fit \nh\ (defined as the equatorial \nh, see Murphy \& Yaqoob 2009)
needs to be higher to reproduce a given spectral shape, with respect to larger $\Theta_{obs}$.
We note also that for heavily CT sources (F$_{PDF}^{CT}=1$) the \nh\ PDF is less affected by these changes.

\section{Bias corrections}

\subsection{Differential Sky coverage}

The sky coverage of COSMOS-Legacy was computed converting count rates into fluxes in different bands,
assuming a power-law spectrum with $\Gamma = 1.4$ and Galactic \nh$=2.6\times10^{20}$ \cm2 (Civano et al. 2016).
This is appropriate if averaging over the intrinsic \nh\ of the whole AGN population, 
and correctly applies, on average, also to the M16 sample of sources with more than 30 net counts in the
0.5-7 keV band.

However, to correcly derive \nh\ distributions we need to take into account the fact that the conversion between detected
counts and emitted flux depends on the spectral shape (see e.g. Vito et al. 2014; Liu et al. 2017).
In the case of our sample of CT AGN,  the source \nh\ and redshift change the conversion factor
between counts and flux in the sense that 30 full band counts for a CT source tipically
correspond to a higher observed 2-10 keV flux with respect to an average power-law.

We therefore recomputed the 2-10 keV sky coverage in the three redshift bins adopted in the {log}N-{log}S analysis
for a spectrum with \nh$=10^{24}$ \cm2 (see Fig.~\ref{fig_skycov}), and applied these corrected sky coverages when deriving 
the source counts for the {log}N-{log}S.
The differences in sky coverage, i.e. the difference of area at a given flux, have a maximum of 30, 10 and 3\% for  z1, z2 and z3 respectively, 
for the faintest fluxes covered by our sample.
The effect of these differences is however very small, appreciably increasing the cumulative number counts of 
CT AGN only in the last two data points of Fig.~\ref{fig:lognlogs} and only in the first redshift bin.
This correction instead does not affect the conclusions of Sec.~5 (Fig.~\ref{fig:ctf})
since, as explained in Sec.~5, we are taking into account only sources well above the survey flux limit.

\begin{figure}
	\includegraphics[width=\columnwidth]{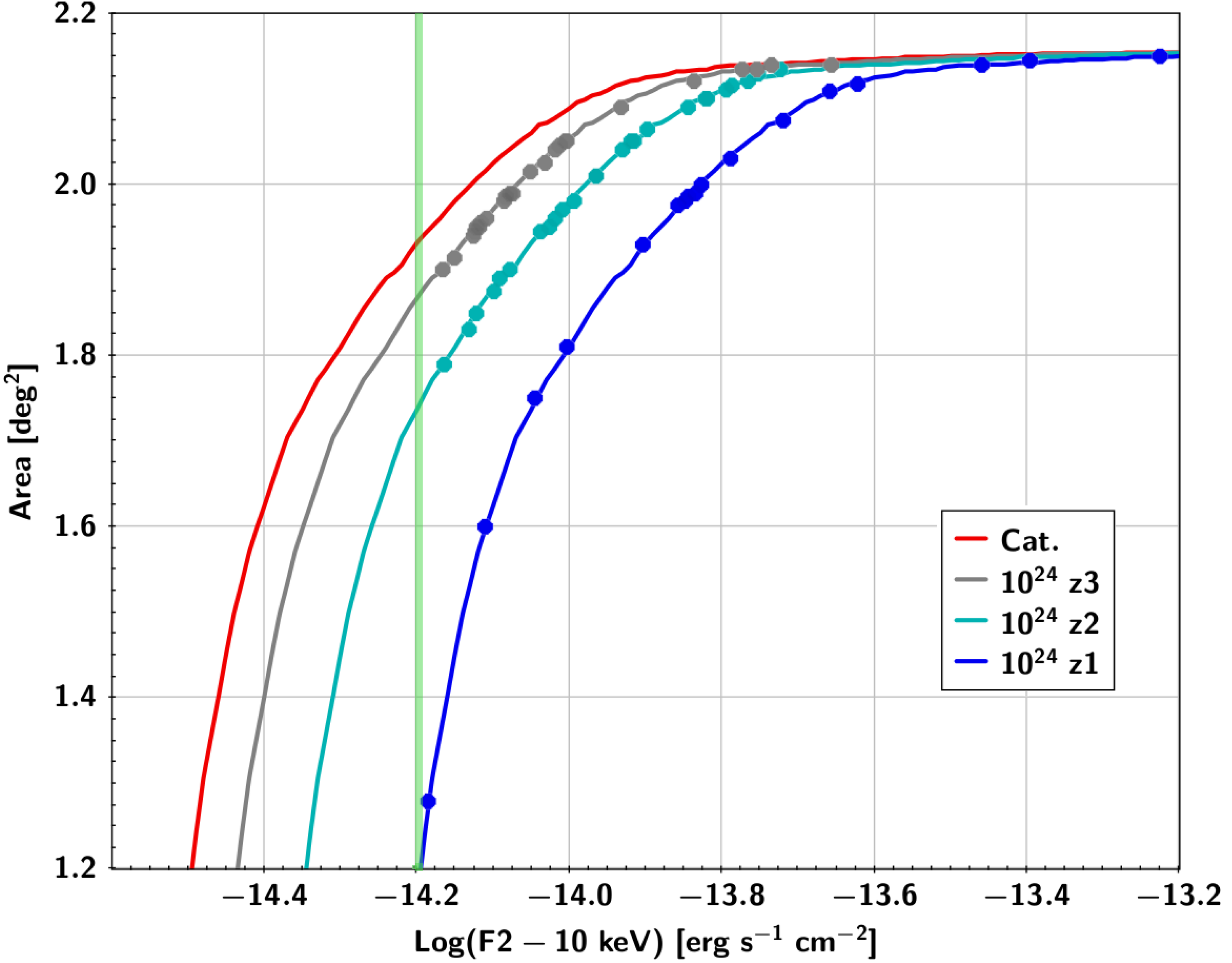}
    \caption{2-10 keV sky coverage of the COSMOS-Legacy survey, as computed for the whole catalog (red curve) and as
    recomputed here for sources with \nh$=10^{24}$ \cm2, in the three redshift bins (blue z1, cyan z2, gray z3). 
    The green vertical line shows the nominal flux limit for the CT sample.}
    \label{fig:skycov}
\end{figure}

\subsection{Classification bias}

We performed simulations to derive the fraction of CT sources correctly identified as such, for a given flux and redshift.
We followed BU12 and simulated $10^3$ spectra in each of the three redshift bins
and five intervals of flux, assuming \nh$=10^{24}$ \cm2 and $\Gamma=1.8$, and a secondary powerlaw with
$f_{scatt.}=3\%$ fixed.
We then performed the same spectral analysis described in Sec~2.2 and analyzed the derived PDF distributions.
Since the input \nh\ for the simulation is centered at $10^{24}$ \cm2, 
we expect the fraction of PDF above this value
to be $\sim0.5$ (the PDF should be symmetric around the input value).
Indeed, this value is recovered for the highest flux bins, while at the faintest fluxes (below $10^{-14}$ \cgs)
the average PDF fraction above $10^{24}$ \cm2\ is significantly lower than the expected 0.5. 
This translates into a fraction of correctly identified CT sources
of $\sim0.6$, 0.8 and 0.95 at z1, z2 and z3 respectively 
(i.e. it is easier to identify a faint CT source if it is located at high redshift, than if it is 
at low z).
We note that sources with \fct\ close to one can have this value $>1$ once the classification bias correction is applied.
This has the meaning of accounting for other sources with similar \nh\ PDF and number of counts that are instead missing
due to the classification bias.
We also note that the choice of limiting the analysis to sources with more than 30 counts, translates into
a less severe correction for misclassification, with respect to BU12, that has no cut in number of counts.

Finally we simulated $10^3$ spectra for {log}(\nh)=21,22,23 (\cm2) in the three redshift bis and five flux intervals.
The contamination, i.e. C-thin sources with a sizable fraction of the \nh\ PDF in the CT regime is negligible
at these flux and counts levels: the average \nh\ PDF of the simulated spectra with input {log}(\nh)=23 (\cm2) at z1 
and the lowest fluxes (below $10^{-14}$ \cgs) has only a fraction of 0.02 
above the CT threshold, while the others have none.

% Don't change these lines
\bsp	% typesetting comment
\label{lastpage}
\end{document}